\documentclass[11pt]{article}

\usepackage{setspace,graphicx,epstopdf,amsmath,amsfonts,amssymb,amsthm,versionPO}
\usepackage{marginnote,datetime,enumitem,rotating,fancyvrb,ulem}
\usepackage{hyperref,float}

\usepackage[longnamesfirst]{natbib}
\usdate

\usepackage{bm}
\usepackage{booktabs}
\usepackage{pdflscape}
\usepackage{chngcntr}
\usepackage{longtable}
\usepackage{subfig}
\usepackage{graphicx}
\usepackage{adjustbox}
\usepackage{multirow}

\usepackage{titlesec}
\newcommand{\periodafter}[1]{#1.}
\titleformat{\subsubsection}[runin]
{\normalfont\bfseries}{\thesubsubsection}{1em}{\periodafter}

\excludeversion{notes}		% Include notes?
\includeversion{links}          % Turn hyperlinks on?

\iflinks{}{\hypersetup{draft=true}}

\ifnotes{%
\usepackage[margin=1in,paperwidth=10in,right=2.5in]{geometry}%
\usepackage[textwidth=1.4in,shadow,colorinlistoftodos]{todonotes}%
}{%
\usepackage[margin=1in]{geometry}%
\usepackage[disable]{todonotes}%
}

\makeatletter\let\chapter\@undefined\makeatother % Undefine \chapter for todonotes

\usepackage{endnotes}    % Use endnotes instead of footnotes
\DeclareMathOperator*{\argmin}{arg\,min}

\newcommand{\abs}[1]{ \left| #1 \right| }

\newcommand{\indicator}[1]{ \mathrm{I} \left( #1 \right)}

\begin{document}
%\begin{CJK}{UTF8}{kai}

\setlist{noitemsep}  % Reduce space between list items (itemize, enumerate, etc.)
\onehalfspacing      % Use 1.5 spacing

\title{\Large\bf Estimating Contagion Mechanism in Global Equity Market with Time-Zone Effect\thanks{Wu, boyao.wu@uibe.edu.cn, School of Banking and Finance, University of International Business and Economics; Huang (corresponding author), difang@hku.hk, Faculty of Business and Economics, University of Hong Kong; Chen, zizizhuzhu0312@163.com, School of Management Science and Engineering, Central University of Finance and Economics. We are grateful to the anonymous referee, Bing Han (the editor), Monica Billio, Christian Brownlees, Jiti Gao, Kewei Hou, Bonsoo Koo, Oliver Linton, Mila Getmansky Sherman and the audiences at various seminars and universities for their constructive comments. Boyao Wu gratefully acknowledges the financial support from the National Natural Science Foundation of China (Grant Number 72271055) and China Ministry of Education Humanities and Social Sciences Youth Foundation (Grant Number 22YJC790194). Boyao Wu is immensely grateful to Miss Yimeng Li for her invaluable contribution in meticulously plotting Figure 1, as well as her unwavering support and critical assistance throughout the paper. Difang Huang gratefully acknowledges the financial support from the Seed Fund for Basic Research from the University of Hong Kong (Grant Number 2201100299) and Basic Research Fund from HKU Education Consulting (Shenzhen) Co, LTD. (Grant Number SZRI2023-BRF-11). Muzi Chen gratefully acknowledges the financial support from the National Natural Science Foundation of China (Grant Number 71673315, 71850008, 72003110). All errors are our own.}}

\author{{\bf Boyao Wu} \\
University of International Business and Economics \vspace{0.2cm} \\
{\bf Difang Huang} \\
The University of Hong Kong \vspace{0.2cm} \\
{\bf Muzi Chen} \\
Central University of Finance and Economics  \\
}

\date{\today}

\maketitle
\thispagestyle{empty}

\clearpage
\pagenumbering{arabic}

\doublespacing
% Use endnotes instead of footnotes - redefine \footnote command
%\renewcommand{\footnote}{\endnote}  % Endnotes instead of footnotes

\begin{center}{\Large\bf Estimating Contagion Mechanism in Global Equity Market with Time-Zone Effect}\end{center}

\vspace*{1in}

\centerline{\bf Abstract}
\medskip

This paper proposes a time-zone vector autoregression (VAR) model to investigate comovements in the global financial market. Analyzing daily data from 36 national equity markets, we explore the subprime and European debt crises using static analysis and the COVID-19 crisis through a rolling window method. Our study of comovements using VAR coefficients reveals a resonance effect in the global system. Findings on densities and assortativities suggest the existence of the transmission mechanism in all periods and abnormal structural changes during the crises. Strength analysis uncovers the information transmission mechanism across continents over normal and turmoil periods and emphasizes specific stock markets' unique roles. We examine dynamic continent strengths to demonstrate the contagion mechanism in the global equity market over an extended period. Incorporating the time-zone effect significantly enhances the VAR model's interpretability. Signed networks provide more information on global equity markets and better identifies critical contagion patterns than unsigned networks.



\clearpage

%===================================================================================%
%===================================================================================%
%===================================================================================%
%===================================================================================%
%===================================================================================%
%===================================================================================%
%===================================================================================%
%===================================================================================%
%===================================================================================%
%===================================================================================%

\section{Introduction}

\noindent
Increased economic globalization continues to intensify the connections between countries, especially in the financial field \citep{acemoglu2015systemic, summer2013financial}. The flows of capital and labor in the global market give rise to close interactions between countries and lead to greater economic integration around the world \citep{elliott2014financial, glasserman2016contagion}. Cross-border investments and international trade promote the mutual penetration of the global financial system, evidenced by comovements in stock markets. Such comovements may also act as transmission channels through which financial risk spreads beyond borders, triggering a global chain reaction that amplifies the damage caused by investor panic and market illiquidity in one location \citep{bernanke2004what, fratzscher2019monetary}.

It is crucial for economists, policy-makers, and practitioners to understand comovements in global financial markets and the underlying contagion mechanisms. This paper proposes a time-zone vector auto-regressive (VAR) model with LASSO to characterize daily comovements across stock markets. Due to differences in time zones, stock markets in Asia stop trading before the European markets open, and the European markets close within a few hours of American markets opening. Therefore, fluctuations in American markets also partly reflect information in other markets overseas, suggesting the direct application of the VAR model to daily data may over-emphasize the importance of some American markets and lead to biased interpretations.

Our model provides a new framework to understand comovements between national stock markets. We propose the connection effect, which refers to the existence of comovements, and the resonance effect, which refers to the extent to which agents interact with others in the same direction, to describe the connectedness of the global financial markets. Such synchronous vibrations are revealed in whether comovements exist in related markets and whether these comovements share the same direction (i.e., national stock indixes rise or fall simultaneously). During periods of economic expansion, stock markets in many countries show upward trends (same-direction comovements), accumulating bubbles to some degree \citep{han_yang_2013}. Moreover, during periods of economic recession, a wide range of declines can be observed in many national markets (same-direction comovements), accelerating the spread of financial risks and contributing to more significant losses worldwide.

We apply the time-zone VAR model to investigate the subprime and European debt crises using the daily returns from 36 national equity markets. Based on unsigned networks and positive and negative subnetworks, we use the density, continent assortativities, and degree assortativities to study the macrostructure of the global equity market. Density results suggest structural changes of the global market in crisis periods, while assortativity results demonstrate a persistent transmission mechanism existing in the global market over both normal and turmoil periods. Analyses on continent degrees and strengths further reveal this transmission mechanism mainly reflected as the information flow across continents (revealed in same-direction comovements), self-regulatory interactions within continents (revealed in opposite-direction comovements), and the spillover effect in crisis periods. Our results are consistent with the literature on the roles of demand shocks \citep{forbes2002no, forbes2004asian}, cash-flow shock \citep{bekaert2014global, boyer2006how}, and currency channel during these crises
\citep[see][for detailed explanations]{albuquerque2009global, hau2017the}. We also identify the roles of national equity markets and confirm the key impact of the United States market on the global stock system over different crisis periods, in line with the role of information frictions on investors that cause the under-reaction of equity prices in certain countries \citep{hong1999unified, hong2007industries}.

We further apply our model to study the COVID-19 crisis by presenting the dynamic strengths of continents and demonstrating the contagion mechanism in the global equity market using the rolling-window estimation. Dynamic analysis indicates the global market has exhibited abnormal behaviors and opened up new channels for information transmission beyond the traditional contagion mechanism. Furthermore, we show that the impacts of the COVID-19 crisis are more sudden and severe than those of the subprime and European debt crises, in line with the evidence in a series of influential papers that COVID-19 acts more like a short-term shock rather than a long-term crisis \citep{caballero2021model, duchin2021covid, barry2022corporate, guerrieri2022macroeconomic, eichenbaum2021macroeconomics, acemoglu2021optimal, baqaee2022supply, augustin2022sickness}.

By comparing the VAR model with and without explicitly taking the time-zone effect into account, we find that incorporating the time-zone effect significantly increases the in-sample performance of the VAR model. This finding highlights the importance of considering the time-zone effect when examining the transmission mechanisms of the daily global stock market. Moreover, incorporating the time-zone effect allows for a more accurate representation of the interconnectedness between different stock markets, which is crucial in understanding the dynamics of global financial markets.

It is important to note that taking comovement directions (reflected as VAR coefficient signs or link signs) into account can help scholars better understand the structural changes of the global equity market in different periods. For instance, during periods of crisis, stock prices may experience massive declines \citep{calomiris2012stock}, while in other periods, an active self-regulatory mechanism in some national stock markets may mitigate the impact of the crisis \citep{almeida2012corporate}. By telling same-direction and opposite-direction comovements, researchers can gain a deeper understanding of the global equity market's behavior, which in turn can lead to improved policy decisions and more effective risk management strategies.

Our paper contributes mainly to two strands of literature. First, our work contributes to the growing literature on the characterization of comovements and contagion mechanisms on the dynamic evolution of the global network \citep{diebold2012better, han2022social}. Some ascribe comovements to interconnections of economic fundamentals among countries and the arbitrage of investor behavior once deploying international investment portfolios \citep{bodnar2002pass, boyd2005the}. Others argue that the contagion between markets results from attempts by investors who utilize other markets’ fluctuations to gain their own market information \citep{andersen2006real, Chen2021entropy, CHEN2022, Huang2022}. Our paper contributes to this strand of literature with a detailed analysis of the sources of contagion by considering comovement directions, allowing us to disentangle economic mechanisms regarding contagion. To the best of our knowledge, existing literature discusses the financial contagion by considering the existence of comovements based on the strengths and directions of links, without taking into account the comovement directions reflected as coefficient signs in the VAR model or equivalently link signs in the network, which may create a significant loss of information \citep[e.g.,][]{diebold2009measuring, demirer2018estimating, claeys2014measuring, de2013bank}. As the positive coefficients in VAR models are initially different from negative ones, we define the resonance effect in the global equity market and provide a new network perspective to better understand comovements in the global stock market. Our empirical results suggest that identifying comovement directions is beneficial to discovering more information about national stock markets and identifying critical contagion mechanisms patterns.

Second, there is a vast literature on the crisis, including subprime, European debt, and COVID-19 crises \citep{almeida2012corporate, Bao2021JFQA, barigozzi2019nets, calomiris2012stock, diebold2009measuring, eichengreen2012how, frankel2010are, hau2017the, li2023impact}. For instance, \citet{billio2012econometric} use the Granger causality test to evaluate before-after relationships among financial firms while \citet{billio2018bayesian} propose a Bayesian Markov-switching regression model for tensors to study time-varying sparsity patterns of the network structure during the subprime and European debt crises. \citet{geraci2018measuring} adopt the time-varying parameter VAR model to estimate the dynamic network in the equity market and show the decrease in network connectivity during the European debt crisis. \cite{monica2021covid19} propose a Markov-switching graphical model to investigate changes in systemic risk for the COVID-19 crisis. Our paper contributes to this strand of literature by using daily data and taking the time-zone effect into account to model the daily global stock market. The fluctuations of stock markets in the Americas partly reflect information from the global market, implying the direct application of the VAR model to daily data may exaggerate the importance of some equity markets in the Americas and cause biased interpretations \citep{rapach2013international}. The time-zone effect is vital to understanding the transmission mechanisms of other stock markets during the crises.

The remainder of our paper is organized as follows. Section~\ref{section: methodology} discusses model specification and the econometric methodology, Section~\ref{section: data} shows data, Section~\ref{Section Static Estimation} and~\ref{Section: Dynamic Estimation} presents the static and dynamic estimation results, Section~\ref{Section: Additional Results} describes the additional results, and Section~\ref{section: conclusion} concludes the paper.

%===================================================================================%
%===================================================================================%
%===================================================================================%
%===================================================================================%
%===================================================================================%
%===================================================================================%
%===================================================================================%
%===================================================================================%
%===================================================================================%
%===================================================================================%

\section{Methodology}
\label{section: methodology}

\subsection{Time-Zone VAR Models with LASSO}

The VAR model plays an important role in understanding the global economic system \citep{dees2007exploring, baker2009global, pukthuanthong2009global}. However, ignoring the time-zone effect may lead to the problematic application of VAR models in characterizing the daily global stock market, as significantly distinct trading windows exist among national equity markets. On a typical day, traders from American markets, including United States and Canada, know that day's trading in Asia due to time-zone differences and receive the information from European stock exchanges. Compared with stock markets in other continents, American equity markets encompass more information and, hence, place more weight on the daily global stock market \citep[e.g.,][]{bekaert2009international,tong2011the,Yu2021CSU}. Consequently, the time-zone effect caused by geographic location profoundly influences the global equity market, especially when studies involve daily trade data.

Table \ref{Tab National Indexes} reports the market opening and closing times for 36 national equity markets. Typically, the Asian markets close at approximately 1:00 am Eastern Standard Time (EST) (with the exceptions of Indonesia, Malaysia, and Thailand, which close at 4:00 am) before any European or American markets open. Meanwhile, European trading commences at 3:00 am EST, while the American markets begin trading before 9:30 am EST. The European markets subsequently close at about 11:30 am (except for Germany closing at 2:00 pm EST), followed by American markets at about 4:00 pm EST. Given the time-zone effect in market opening and closing times, all of the information released before one trading day cannot be incorporated into all equity markets on that same day.\footnote{\citet{rapach2013international} address a similar problem of international monthly stock return predictability by excluding the last trading day of the month $t$ when computing the monthly stock return if the close of the equity market in a country occurs before the United States market closing.}

\bigskip
\centerline{\bf [Place Table~\ref{Tab National Indexes} about here]}
\bigskip

We propose a time-zone VAR model by involving the time-zone effect to investigate interactions among the daily global stock market. The VAR specification is a multi-variate method to quantify the relationships among countries in the global equity market.\footnote{By contrast, pairwise measurements, such as the Granger causality, may not be appropriate for describing sophisticated linkages in complex systems from the theoretical view. For any given agent pair, the white noise assumption in the Granger causality test implies that there are no connections between the two concerned agents and the rest. In other words, these two agents are presumed to be isolated from the entire system, which may contradict the networking structure \citep{bekaert2009international,bekaert2021risk,Yu2023,YU2023JEF}.} According to geographic location, we divide the global stock market into three subsets--Asia (As), Europe (Eu), and Americas (Am)--such that national equity markets in the same region share very close trade windows. Indeed, there are few overlapped trade hours in national equity markets from different subsets. The time-zone VAR model is denoted as:
\begin{align}
\left\lbrace
\begin{array}{cll}
	r_{i_0 t} = &\sum\limits_{i \,\in\, As} \beta_{i_0 i} r_{i, t-1} + \sum\limits_{j \,\in\, Eu} \beta_{i_0 j} r_{j, t-1} + \sum\limits_{k \,\in\, Am} \beta_{i_0 k} r_{k, t-1} + \varepsilon_{i_0 t},  &  i_0 \in As, \\
	r_{j_0 t} = &\sum\limits_{i \,\in\, As} \beta_{j_0 i} \bm{\uline{r_{i, t}}} + \sum\limits_{j \,\in\, Eu} \beta_{j_0 j} r_{j, t-1} + \sum\limits_{k \,\in\, Am} \beta_{j_0 k} r_{k, t-1} + \varepsilon_{j_0 t},  &  j_0 \in Eu, \\
	r_{k_0 t} = &\sum\limits_{i \,\in\, As} \beta_{k_0 i} \bm{\uline{r_{i, t}}} + \sum\limits_{j \,\in\, Eu} \beta_{k_0 j} \bm{\uline{r_{j, t}}} + \sum\limits_{k \,\in\, Am} \beta_{k_0 k} r_{k, t-1} + \varepsilon_{k_0 t},  &  k_0 \in Am,
\end{array}
\right.
\label{Eq Time-Zone VAR Model}
\end{align}
where $ r_{lt} $ is the return of the $ l^{th} $ national equity market, and $ \varepsilon_{lt} $ is the normal white noise. The differences between the time-zone VAR and classic VAR model are underlined and marked in bold, deriving from  the latest information on Asian and European markets. Equation~\eqref{Eq Time-Zone VAR Model} reveals the trading sequence in the global stock market that first opens in Asia and finally closes in the Americas. Specifically, Asian markets start to trade; hence, they are primarily influenced by the global market from the previous day. With only a few overlapping hours between the two kinds of markets, European markets have access to the latest closed prices from Asia, implying that current Asian markets and previous European and American markets influence European stock markets. As the last continent to start trading, American markets receive the most up-to-date information from those closed markets in Asia and Europe, meaning their indexes may respond to their own indexes on the last day and to Asian and Europe indexes from that current day.\footnote{We do not include explanatory variables such as industry or macroeconomic factors in our time-zone VAR model. See details in Appendix \ref{Appendix A Reasons of Excluding Explanatory Variables in Equation 1}.} We show additional technical details on the estimate of the time-zone VAR model in Appendix \ref{Appendix A Methodology}.

The time-zone VAR model~\eqref{Eq Time-Zone VAR Model} only involves the first-order autoregressive term instead of more lagged values. One reason is that the first autoregressive order may contain more information than other lagged terms. For example, lagged United States returns significantly predict returns in numerous non-United States industrialized countries \citep{rapach2013international}. The other reason is that involving more lagged terms marginally improves the interpretation of model~\eqref{Eq Time-Zone VAR Model} but dramatically increases the number of unknown VAR coefficients and undermines the reliability of the used model, especially using limited data.\footnote{The time-zone VAR model assumes the comovements, reflected as VAR coefficients, are stable during a certain sample period. Our model also allows the time-varying comovements between different countries from dynamic views based on rolling window estimation. Relative results are presented in Section \ref{Section: Dynamic Estimation}.}

%===================================================================================%
%===================================================================================%
%===================================================================================%
%===================================================================================%
%===================================================================================%

\subsection{Connection and Resonance Effect}

Identifying contagion is a big issue that has sparked continued discussion over the past two decades, with many scholars attempting to understand this phenomenon from distinct aspects. Instead of distinguishing subtle differences among these works, we consider a general definition of contagion that covers various theoretical models \citep{claeys2014measuring,karolyi1996markets, forbes2002no, bae2003new, jondeau2018moment}. We view financial contagion as a sudden shock that arises from one agent to the financial system, whose transmission cannot be explained by economic fundamentals.\footnote{Specifically, contagion is different from normal interdependence as comovements between agents that cannot be ascribed to fundamental linkages \citep{kaminsky2000crises} or as excessive comovements across agents \citep{forbes2002no}.} Under the networking framework, links between agents represent comovements between national equity markets, and coefficients in the time-zone VAR model \eqref{Eq Time-Zone VAR Model} characterize the transmission mechanism. To the best of our knowledge, existing literature studies the financial contagion by considering the existence of comovements based on the strengths and directions of links \citep[e.g.,][]{diebold2009measuring, demirer2018estimating, de2013bank}. However, only a few papers consider the comovement directions reflected in link signs, which is a significant loss of information.

We introduce link signs into the analysis based on the strengths and directions of connections and propose the concepts of `connection effect' and `resonance effect' to shed light on studying financial contagion from a networking perspective. The connection effect refers to the extent to which national equity markets interact with others without considering link signs, reflecting comovements in the system. The resonance effect refers to the extent to which agents build positive connections with others (i.e., the degree of connectedness in the positive subnetwork with the adjacency matrix $\bm{A}^+ = (|\bm{A}|+\bm{A})/2 $ and the weight matrix $\bm{W}^+ = (|\bm{W}|+\bm{W})/2 $).\footnote{The $ i^{th} $ row, $ j^{th} $ column elements in matrix $ |\bm{A}| $ and $ |\bm{W}| $ are defined as $ |a_{ij}| $ and $ |w_{ij}| $, respectively.} The resonance effect is reflected in positive links among agents that correspond to those same-direction comovements between national equity markets. In expansion periods, the synchronous growth in national equity indexes results from favorable economic conditions and positive expectations on the global market, where the resonance effect acts as the channel through which to transmit information. In the recession periods, same-direction comovements (the resonance effect) are reflected in widespread declines in national indexes, where the resonance effect accelerates the contagion spread and causes greater losses.\footnote{The resonance effect originates from physics, describing that the significantly increased amplitude occurs when a system oscillates at a specific frequency. This phenomenon happens when the system transfers energy between distinct storage modes and may give rise to the collapse of the whole system.} Also, Figures \ref{Fig HeatMapVARCoefs_SubprimeBefore} to \ref{Fig HeatMapVARCoefs_EuDebtAfter} in Appendix \ref{Appendix VAR Coefficients and Link Signs} present all link signs in generated global networks over different sample periods in the form of heatmaps.

%===================================================================================%
%===================================================================================%
%===================================================================================%
%===================================================================================%
%===================================================================================%

\subsection{Network Properties}

In social networks, agents (nodes) form links to communicate information to other agents, and links act as information channels to determine network features and agent statuses. We use the densities and assortative coefficients to investigate the structural changes in the global stock network over different periods (see details in Sections \ref{Section Density Analysis} and \ref{Section Mixing Patterns}). The two measurements, degrees and strengths, are used to discover the information transmission mechanism within and between continents and further to identify the statuses of national equity markets in different periods (see details in Sections \ref{Section Continent Analysis Based on Degrees} and \ref{Section Country Status Analysis Based on Strengths}). Let $ a_{ij} $ and $ w_{ij} $ be the $ i^{th} $ row and $ j^{th} $ column elements in the adjacency matrix $ \bm{A} $ and the weight matrix $ \bm{W} $, respectively. Moreover, the adjacency matrixes of unsigned network, positive and negative subnetworks are defined as:
\begin{align}
	\bm{A}^U = \abs{\bm{A}}, \qquad  \bm{A}^+ = \frac{\abs{\bm{A}} + \bm{A}}{2}, \qquad  \bm{A}^- = \frac{\abs{\bm{A}} - \bm{A}}{2},
\label{Eq Adjacency Matrix of Unsigned Positive Negative Networks}
\end{align}
respectively, where positive and negative subnetworks are parts of the original one $ \bm{A} $ consisting of only positive and negative links.

%===================================================================================%
%===================================================================================%
%===================================================================================%

\subsubsection{Density}

The density describes how national equity markets interact with others in networks. Densities of the unsigned network, positive and negative subnetworks are defined as
\begin{align}
&\text{Density of Unsigned Network} = \frac{1}{N(N-1)} \sum_{i,j = 1}^{N} \abs{a_{ij}},
\label{Eq Density of Unsigned Network}   \\
&\text{Density of Positive Subnetwork} = \frac{1}{N(N-1)} \sum_{i,j = 1}^{N} \abs{a_{ij}} \indicator{ a_{ij} > 0 },
\label{Eq Density of Positive Subnetwork}   \\
&\text{Density of Negative Subnetwork} = \frac{1}{N(N-1)} \sum_{i,j = 1}^{N} \abs{a_{ij}} \indicator{ a_{ij} < 0 },
\label{Eq Density of Negative Subnetwork}
\end{align}
respectively, where $ \indicator{\cdot} $ is the indicator function.

%===================================================================================%
%===================================================================================%
%===================================================================================%

\subsubsection{Assortativity}

The connection patterns in networks have a profound impact on diffusion behaviors like information transmission and risk contagion. The assortativity is important to help us better understand those connection patterns by unraveling who tends to be connected to whom in networks \citep{jackson2008social}. Specifically, the assortative mixing pattern (i.e., the positive assortativity) depicts the propensity that agents to connect with others with similar features \citep{newman2002assortative, newman2003mixing}, implying the homophily effect well studied in sociology \citep{mcpherson2001birds}, economics \citep{currarini2009economic} and finance \citep{stolper2019birds}. By contrast, the negative assortativity implies the opposite topological pattern: dissimilar agents are more likely to build connections. We discuss two types of assortative coefficients based on continents and degrees.

The continent assortativity is defined as:
\begin{align}
\mathrm{Continent \; Assortativity} = \frac{ \sum_{k=1}^{n} e_{kk} - \sum_{k = 1}^n e_{k \cdot} e_{\cdot k} }{ 1 - \sum_{k = 1}^n e_{k \cdot} e_{\cdot k} },
\label{Eq Assortatative for Continents}
\end{align}
where $ e_{ij} $ is the fraction of edges in a network connecting from continent $ i $ to continent $ j $, $ e_{i \cdot} = \sum_{k=1}^{n} e_{ik} $, $ e_{\cdot j} = \sum_{k=1}^{n} e_{kj} $ and $ n $ is the number of continents. The positive value of the continent assortativity implies that nodes tend to build relations with others in the same continent, while the negative value denotes the opposite situation. The degree assortativity is defined as:
\begin{align}
\mathrm{Degree \; Assortativity} = \frac{\sum_{ij} ij(d_{ij} - q_i^{out} q_j^{in})}{\sigma_{in} \sigma_{out}},
\label{Eq Assortative for Degrees}
\end{align}
where $ d_{ij} $ is the fraction of edges in a network connecting from nodes with degree $ i $ to nodes with degree $ j $, $ q_i^{out} = \sum_{k=1}^n d_{ik}, $ $ q_j^{in} = \sum_{k=1}^n d_{kj} $, $ \sigma_{in} $ and $ \sigma_{out} $ are standard deviations of the indegree and outdegree distributions. The positive degree assortativity implies that high-degree (low-degree) agents tend to associate with other high-degree (low-degree) agents, and two types of communities with high- and low-degree agents emerge in networks such that each community has a higher average degree than the whole network \citep{silva2016network}. Conversely, in the negative degree-assortative network, high-degree agents are prone to connect with those low-degree ones, suggesting the existence of a `core-periphery' topology in networks.

Widely existing in financial systems, the core-periphery structure attracts much attention because of its implication for higher systemic risks \citep[e.g.,][]{elliott2014financial, jackson2021systemic, lee2013systemic, silva2016machine}, making the system less resilient to financial crises \citep{newman2003mixing}. In a core-periphery network, peripheral agents mainly establish connections with core agents but not among similar peers, while core agents act as intermediaries to connect the remainder of the network. In fact, the assortative mixing pattern is related to the core-periphery structure \citep{jackson2008social}, and \citet{silva2016machine} provide evidence that the degree assortativity is a good proxy to measure how compliant a network is to the core-periphery structure. Also, it is common to find in empirical studies on financial networks that the negative degree assortativity is related to the core-periphery structure \citep{soramaki2007topology, huser2015too, silva2016network, silva2016financial, iori2008network, INTVELD201427, roukny2014network, siudak2022network, bargigli2015multiplex, berndsen2018financial, schiavo2010international}. Given the above reasons, we discuss the degree assortativity to investigate whether the core-periphery pattern exists in the global equity network.

%%\footnote{
%This implies that the targeted strategy is an effective precaution in this situation.
%
%Due to the limited resources require the adoption of an appropriate policy for risk management, one of the possible strategies, targeted strategy, deletes a few agents from the risk contagion graph in the context of networks and frees a small portion of agents from risks \citep{newman2003mixing}.
%}

%===================================================================================%
%===================================================================================%
%===================================================================================%

\subsubsection{Strength}

\begin{figure} 
\begin{center}
	\includegraphics[width=\textwidth]{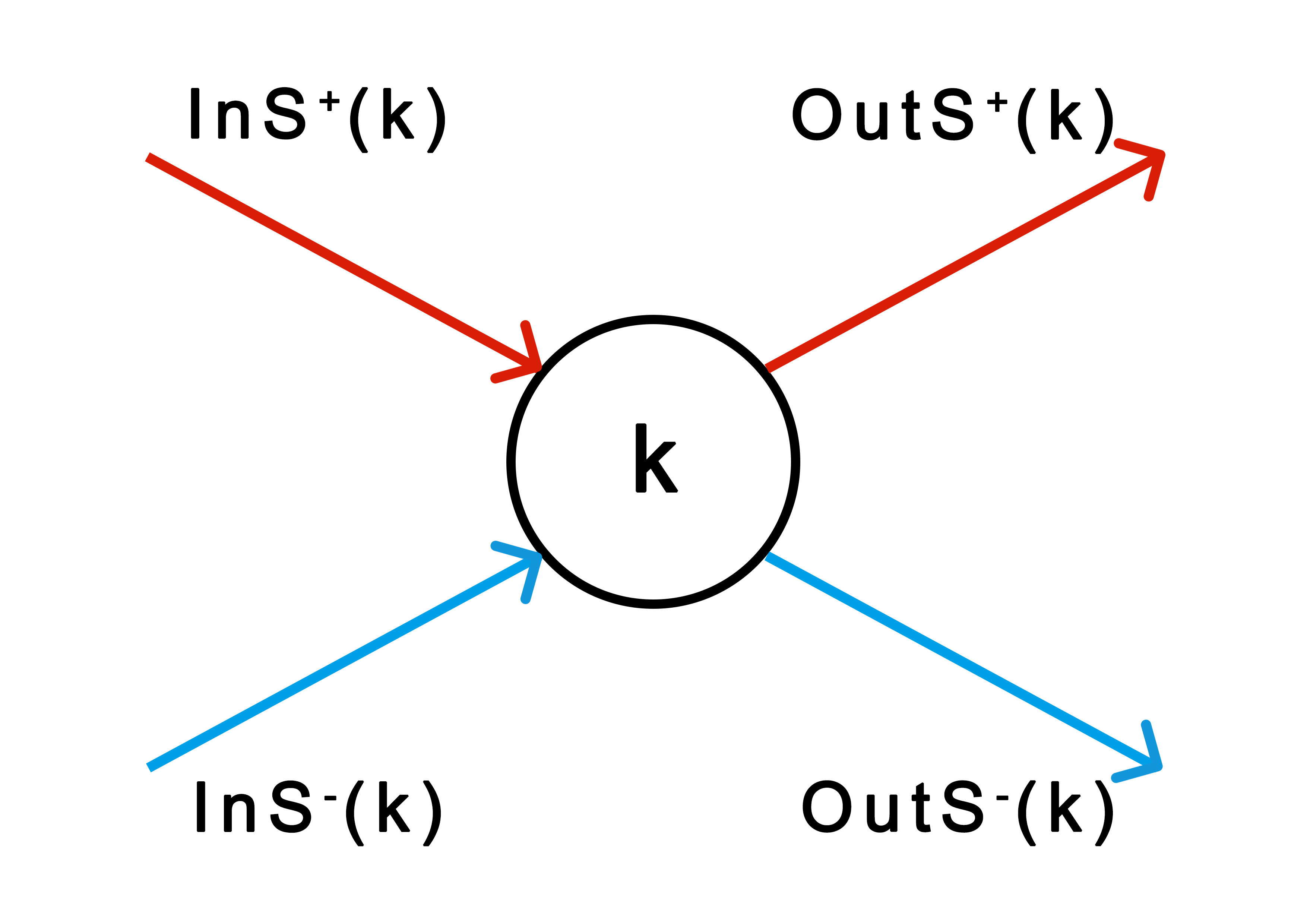}
	
	\caption{\textbf{Four types of interactions between agent $ k $ and the system.} Red and blue edges represent positive and negative links in the adjacency matrix. Interaction measurements are defined in Equations \eqref{Eq Strengths for Positive Weights} and \eqref{Eq Strengths for Negative Weights}.}
	
	\label{Fig Four Types of Interactions}
\end{center}
\end{figure}

As \citet{barrat2004architecture} suggested, networks display significant heterogeneity in the number of links and the intensity of connections, making it difficult to identify national equity markets' roles in the global market over different periods. To fully exploit the valuable information in signs and weights, we use the weight matrixes of positive and negative subnetworks to analyze strengths, defined in Equation~\eqref{Eq Weight Matrices Signed}, respectively, as
\begin{align}
\bm{W}^+ = \frac{|\bm{W}| + \bm{W}}{2}, \qquad  \bm{W}^- = \frac{|\bm{W}| - \bm{W}}{2}.
\label{Eq Weight Matrices Signed}
\end{align}
The in-strengths and out-strengths \citep{barrat2004architecture} for the $ k^{th} $ agent in $ \bm{W}^+ $ and $ \bm{W}^- $ are defined as
\begin{align}
InS^+(k) = \sum_{i=1}^N w_{ik}^+, \qquad OutS^+(k) = \sum_{j=1}^N w_{kj}^+
\label{Eq Strengths for Positive Weights},   \\
InS^-(k) = \sum_{i=1}^N w_{ik}^-, \qquad OutS^-(k) = \sum_{j=1}^N w_{kj}^-
\label{Eq Strengths for Negative Weights},   
\end{align}
where $ w_{ij}^+ $ and $ w_{ij}^- $ are the $ i^{th} $ row, $ j^{th} $ column elements in $ \bm{W}^+ $ and $ \bm{W}^- $, respectively. As suggested in Figure \ref{Fig Four Types of Interactions}, $ InS^+(k) $ and $ InS^-(k) $ reflect two different channels through which the system affects agent $ k $, while $ OutS^+(k) $ and $ OutS^-(k) $ mirror two types of influence from agent $ k $ on the system. The strengths from continent $ \bm{\varPhi} \in \{ \text{As, Eu, Am} \} $ to continent $ \bm{\varPsi} \in \{ \text{As, Eu, Am} \} $ are defined as 
\begin{align}
&\text{Unsigned Strength from Continent } \bm{\varPhi} \text{ to Continent } \bm{\varPsi} %
= \sum_{i \in \bm{\varPhi}} \sum_{j \in \bm{\varPhi}} \left( w_{ij}^+ + w_{ij}^- \right),
\label{Eq Continent Strength Unsigned}   \\
&\text{Positive Strength from Continent } \bm{\varPhi} \text{ to Continent } \bm{\varPsi} %
= \sum_{i \in \bm{\varPhi}} \sum_{j \in \bm{\varPhi}} w_{ij}^+,
\label{Eq Continent Strength Positive}   \\
&\text{Negative Strength from Continent } \bm{\varPhi} \text{ to Continent } \bm{\varPsi} %
= \sum_{i \in \bm{\varPhi}} \sum_{j \in \bm{\varPhi}} w_{ij}^-.
\label{Eq Continent Strength Negative}
\end{align}
Moreover, the degrees from continent $ \bm{\varPhi} \in \{ \text{As, Eu, Am} \} $ to continent $ \bm{\varPsi} \in \{ \text{As, Eu, Am} \} $ can be similarly defined by replacing $ w_{ij}^+ $ and $ w_{ij}^- $ in Equations \eqref{Eq Continent Strength Unsigned} to \eqref{Eq Continent Strength Negative} with $ a_{ij}^+ $ and $ a_{ij}^- $, respectively.

\bigskip
\centerline{\bf [Place Figure~\ref{Fig Four Types of Interactions} about here]}
\bigskip

Next, we consider the net interactive patterns between the given agent $ k $ and the whole system, measured by net in-strengths and net out-strengths as
\begin{align}
\mathrm{NetInS} (k)  = \big| InS^+ (k)  \big| - \big| InS^- (k)  \big|, \quad
\mathrm{NetOutS} (k) = \big| OutS^+ (k) \big| - \big| OutS^- (k) \big|,
\label{Eq Net In- and Out- Strenghts}
\end{align}
where the positive (negative) value of $ \mathrm{NetInS} (k) $ and $ \mathrm{NetOutS} (k) $ suggests the homogenous (heterogeneous) interaction between agent $ k $ and the system, which corresponds to comovements in the same (opposite) direction.

%===================================================================================%
%===================================================================================%
%===================================================================================%
%===================================================================================%
%===================================================================================%
%===================================================================================%
%===================================================================================%
%===================================================================================%
%===================================================================================%
%===================================================================================%

\section{Data}
\label{section: data}

We study the returns of equity market indexes from 36 countries and regions. The sample includes 11 Asian countries and regions (Australia, Malaysia, Indonesia, Korea, Japan, Singapore, New Zealand, Philippines, Thailand, China, and Hong Kong), 19 European countries (Netherlands, Greece, Belgium, France, Germany, Finland, Spain, Ireland, Italy, Denmark, Norway, Sweden, Portugal, Russia, Switzerland, United Kingdom, Poland, Turkey, and Austria) and 6 American countries (Brazil, Chile, Argentina, Mexico, Canada, and United States).\footnote{See Table \ref{Tab National Indexes} for a list of national equity market index codes and opening/closing times.}

We obtain the daily stock return data from the Global Financial Data, spanning from 1 August 2006 to 31 December 2015, to study the influence of the subprime crisis and the European debt crisis.\footnote{Following \cite{ang2007stock} and \cite{rapach2013international}, we obtain the returns of national equity market indexes from the Total Return indexes---Stocks series in Global Financial Data's Total Return Database.} We further divide the sample period into five-time intervals: (1) before the subprime crisis (1 August 2006 to 31 July 2007), (2) during the subprime crisis (1 August 2007 to 31 March 2009), and (3) after the subprime crisis (1 April 2009 to 30 November 2009), (4) during the European debt crisis (1 December 2009 to 16 December 2013), and (5) after the European debt crisis (17 December 2013 to 31 December 2015).\footnote{We determine the intervals using factors including landmark events, equity markets' volatilities, and economic growth. See the details in Appendix~\ref{Appendix A sample divisions}.}

\section{Static Estimation of the Global Equity Market in Subprime and European Debt Crisis}
\label{Section Static Estimation}

In this section, we study the network structural changes in the global equity market and analyze continent degrees and strengths to show the transmission mechanisms in different periods. We also identify the statuses of national equity markets and confirm the notable impact of the United States market on the global stock system over different periods.\footnote{It is important to highlight that we do not apply the static estimation approach to study the global equity market in the COVID-19 crisis because of three reasons: different COVID-19 severity, short duration of COVID-19, and non-stationary exogenous variables on COVID-19 infections. See more discussions on these three reasons in Appendix \ref{Appendix A Not consider static estimation for COVID-19 crisis}.}
%We present the advantages of the time-zone VAR model and improved CV procedure in Sections \ref{Section Comparison between Time-Zone and Classic VAR Models} and \ref{Section Comparisons on Network Structures}, respectively. Section \ref{Section Static Estimation} discusses the network topologies.

%===================================================================================%
%===================================================================================%
%===================================================================================%
%===================================================================================%
%===================================================================================%

\subsection{Density Analysis}
\label{Section Density Analysis}

Figure \ref{Fig Densities Before, During, After the Crisis} shows the densities of generated unsigned network, positive, and negative subnetworks over five periods. We first focus on the connection effect mirrored by the densities of unsigned networks over different periods. There are $ 0.242 $, $ 0.347 $, and $ 0.229 $ unsigned densities in the generated network before, during, and after the subprime crisis, $ 0.275 $ and $ 0.229 $ unsigned densities in the network during and after the European debt crisis. Such a surge in density during the subprime crisis indicates excessive comovements in the global equity market, implying the subprime crisis triggers abnormal changes in the transmission mechanism, as reflected by the connection effect.

The densities of positive and negative subnetworks mirror the resonance effect and provide more information on the two crises. The positive subnetwork density reaches the maximum level during the subprime and European debt crises at $ 0.222 $ and $ 0.214 $, respectively. These two maximum values indicate that the resonance effect is most significant in crises during which the large-scale downward trend in the global equity market gives rise to excessive comovements in the same direction.

Compared with the positive subnetwork, the negative subnetwork's density experiences a more remarkable growth, peaking at $ 0.125 $ in the subprime crisis and remaining relatively stable at $ 0.065 $ in the following three periods. The sudden increase in negative links means that many comovements in the opposite direction merely appear during the subprime crisis, acting as the self-regulatory mechanism to stabilize the global equity market. Consequently, the most significant connection effect in the subprime crisis derives from the growing resonance effect reflected in positive links and the regulatory roles played by some national markets to reduce systemic risk.

\bigskip
\centerline{\bf [Place Figure~\ref{Fig Densities Before, During, After the Crisis}  about here]}
\bigskip

%===================================================================================%
%===================================================================================%
%===================================================================================%
%===================================================================================%
%===================================================================================%

\subsection{Mixing Patterns}
\label{Section Mixing Patterns}

Table~\ref{Tab Assortativity for Continents and Degrees} reports the continent and degree assortative coefficients for the unsigned network, positive and negative subnetworks defined in Equation \eqref{Eq Adjacency Matrix of Unsigned Positive Negative Networks}.
%\begin{align*}
%\bm{A}^U = \abs{\bm{A}}, \qquad  \bm{A}^+ = \frac{\abs{\bm{A}} + \bm{A}}{2}, \qquad  \bm{A}^- = \frac{\abs{\bm{A}} - \bm{A}}{2},
%\end{align*}
%respectively. Here, positive and negative subnetworks are parts of the original one $ \bm{A} $ consisting of only positive and negative links.
The continent assortativities of the unsigned network $ \bm{A}^U $ and the positive subnetwork $ \bm{A}^+ $ are negative over the entire period. Negative continent assortativities of $ \bm{A}^+ $ imply that these national equity markets are prone to influence those from different continents, consistent with the information flow in the global market.\footnote{See details on the information flow in Section \ref{Section Continent Analysis Based on Degrees}.} Therefore, fluctuations in one continent can quickly pass on to the following continent via positive links, and same-direction comovements in the information flow encourage homogeneous behavior across different continents. Eventually, such topological structures strengthen the resonance effect and facilitate the spread of financial risks across continents. The unsigned network $ \bm{A}^U $ presents similar but less prominent patterns in the continent assortativity, showing that the information flow is the dominant transmission pattern in the daily global stock market.

In the negative subnetwork $ \bm{A}^- $, continent assortativities are positive in all five periods, meaning that national equity markets are likely to develop negative relations with others from the same continent. Further, negative links perform regulatory roles that lead national equity markets in the same continent to behave heterogeneously.\footnote{Arguably, the positive continent assortativities imply that unique topological structures of the global market that is a favorable self-regulatory mechanism to diminish those same-direction comovements within each continent.} It is interesting to note that the negative subnetwork's continent assortativity reaches its lowest point at $ 0.335 $ and $ 0.339 $ in the subprime and European debt crises, respectively. This is because two crises result in the self-regulatory mechanism and the spillover effect such that national markets also develop negative connections, reflected as comovements in opposite directions, with those from different continents in the crisis period. More discussions are presented in Section \ref{Section Continent Analysis Based on Degrees}.

\bigskip
\centerline{\bf [Place Table~\ref{Tab Assortativity for Continents and Degrees} about here]}
\bigskip

The degree assortativities are negative in the unsigned network and positive and negative subnetworks over all five periods. High-degree agents are critical in financial networks in transmitting information and spreading risks \citep{billio2012econometric, demirer2018estimating, geraci2018measuring}. In the unsigned network $ \bm{A}^U $, the negative degree assortativities imply that high-degree national equity markets tend to connect those with relatively low degrees, suggesting the presence of the core-periphery topology. The negative assortativity is not a peculiar characteristic and is in line with similar findings in trading networks \citep{li2015unveiling, kyriakopoulos2009network, tseng2010experimental, jiang2010complex}, domestic financial markets, including United States \citep{soramaki2007topology, siudak2022network}, Mexico \citep{martinez2014empirical}, Italy \citep{iori2008network, bargigli2015multiplex}, Netherlands \citep{INTVELD201427}, Germany \citep{roukny2014network}, Brazil \citep{silva2016network, silva2016financial}, and the international financial market \citep{bech2010topology, schiavo2010international}.

Positive and negative subnetworks provide more details on degree assortativities. In the subnetwork $ \bm{A}^+ $, the negative degree assortativities mean that same-direction comovements are more likely to exist between high- and low-degree national equity markets.
% {\color{red} \sout{As discussed in Sections \ref{Section Continent Analysis Based on Degrees} and \ref{Section Country Status Analysis Based on Strengths}, the negative degree assortativity of $ \bm{A}^+ $ is due to the intermediary roles of equity markets like Singapore to transmit the information as well as the benchmark effect of those key markets (e.g., United States) dominating the global finance.}}
On the one hand, this topological feature encourages homogeneous behaviors in national markets and strengthens the resonance effect in the global stock market \citep{camanho2022global}. On the other hand, this feature also accelerates the risk spread from key national equity markets to the global market because a shock hitting markets with high positive degrees will be quickly transmitted to the whole system and hence trigger a systemic crisis, supporting the view that the negative degree assortativity leads to a fragile financial system \citep[e.g.,][]{elliott2014financial, jackson2021systemic, hau2004can, lee2013systemic, silva2016financial, newman2003mixing}.

In the subnetwork $ \bm{A}^- $, the positive continent and negative degree assortativities reveal that high-degree national markets are prone to develop negative links with others from the same continent.
% {\color{red} \sout{This is because the significant variation in liquidity across exchange rates, substantial illiquidity costs, and strong commonality in liquidity across high-degree national markets in portfolio balancing, consistent with the liquidity ﬂow in the international market} \citep{mancini2013liquidity}.}
Consequently, the self-regulatory mechanism mainly works at the continent level but fails to protect the system from the large-scale resonance effect (i.e., massive comovements in the same direction).

%===================================================================================%
%===================================================================================%
%===================================================================================%
%===================================================================================%
%===================================================================================%

\subsection{Continent Analysis}
\label{Section Continent Analysis Based on Degrees}

We consider continent degrees and strengths in the unsigned network and both positive and negative subnetworks, based on adjacency and weight matrixes, respectively, to analyze interactive patterns within and between continents and investigate the connection and resonance effect in the global market.

Throughout the entire period, an apparent `information flow' exists in positive subnetworks: that is, national equity markets from Asia mainly have positive effects on European markets; European markets then positively affect American markets; these American markets positively influence the following day's stock markets in Asia. Table \ref{Tab Continent Degrees Analysis} demonstrates that on average, about $ 86\% $ of positive links flow from Asia to Europe, around $ 59\% $ of positive edges flow from Europe to Americas, and $ 68\% $ of positive connections flow from Americas to Asia.\footnote{Take $ 86\% $ positive links as an example to illustrate how to compute these three percentages. There are $ 106 $ of $ 117 $, $ 127 $ of $ 147 $, $ 80 $ of $ 93 $ positive links from Asia to Europe before, during and after the subprime crisis, and $ 114 $ of $ 137 $ and $ 115 $ of $ 138 $ positive links during and after the European debt crisis. Hence, over five periods, an average of about $ 86\% $ positive links stem from Asia to Europe.} Also, this phenomenon can be notably found in Table \ref{Tab Continent Strength Analysis}, where connections in the information flow account for about $ 85\% $, $ 73\% $, and $ 76\% $ of the total positive weights before, during, and after the subprime crisis, and around $ 83\% $ of the total weights during and after the European debt crisis.\footnote{Take $ 73\% $ of the total positive weights as an example to illustrate how to calculate these four percentages. During the subprime crisis, positive weights of links in the information flow are $ 12.765 $ (As to Eu), $ 4.331 $ (Eu to Am), and $ 5.149 $	(Am to As), and hence the information flow's link weights account for about $ 73\% $ of the total weights ($ 30.401 $).} This interaction pattern indicates that the information flow is a primary transmission mechanism in the global equity market: a continent with earlier closing trade times updates the latest market information and passes it on to the following continent through positive and direct links, where the transmission sequence coincides with the given time zone.

Tables \ref{Tab Continent Degrees Analysis} and \ref{Tab Continent Strength Analysis}  illustrate that most negative relationships exist in the interiors of three continents, and the negative feedback is the dominant interactive pattern within each continent. About $ 98\% $, $ 71\% $, $ 84\% $, $ 71\% $, and $ 81\% $ of negative links exist within continents before, during, and after the subprime crisis, and during and after the European debt crisis, accounting for $ 97\% $, $ 71\% $, $ 88\% $, $ 66\% $, and $ 80\% $ of the total negative weights, respectively. Since most diagonal elements are negative in five periods, national equity markets from the same continent are not likely to rise or fall synchronously (i.e., these markets do not share the same-direction comovement). These interaction patterns are also consistent with the assortativity results in Section \ref{Section Mixing Patterns}, attaching much importance to the financial market stability at the continent level. Besides, more positive and negative links with higher weights stem from American markets to European ones during the subprime and European debt crisis because of the spillover effect.

\bigskip
\centerline{\bf [Place Tables \ref{Tab Continent Degrees Analysis} and \ref{Tab Continent Strength Analysis} about here]}
\bigskip

It is interesting to see that most returns of national indexes lack predictability after controlling for other national markets' returns \citep{bekaert2011what, carrieri2007characterizing}. Only a small part of autoregressive coefficients are significant when imposing equal penalties on coefficients in Equation~\eqref{Eq VAR Models with LASSO-First Country}. According to Table \ref{Tab AR Coefficients} in Appendix \ref{Appendix VAR Coefficients and Link Signs}, less than one-third of national index returns ($ 11.2 $ on average over five periods) at the current moment are helpful to forecast their returns at the next moment. The information flow provides a cogent argument for this phenomenon: for national markets in a given continent, their last continent updates the information on the global market and has better predictability than their autoregressive returns on the previous day.\footnote{Take Asian equity markets as examples. Asia markets directly transmit information to Europe through fluctuations in their returns and indirectly affect the American markets, meaning that American returns reflect information on the Asian markets and the global market. Consequently, for most Asian markets, returns of American stock markets on a given day are more competitive than Asian returns from the previous day when predicting the following day's returns.} On the contrary, the significant autoregressive coefficients reflect some degree of autonomy in their corresponding national markets, resistant to the global market. More importantly, United States is the only national market with significant autoregressive coefficients in five periods, showing its unique role in the global financial system.

It is worth noting that the global equity market experienced abnormal changes in the transmission mechanism during the subprime and European debt crises, reflected in the more significant connection and resonance effect.\footnote{As shown in Tables \ref{Tab Continent Degrees Analysis} and \ref{Tab Continent Strength Analysis}, the total amounts (weights) of positive and negative links maximize at $ 280 (30.401) $ and $ 157 (10.775) $ in the subprime crisis, respectively, exceeding the before period ($ 242 $ positive links with $ 20.137 $ weight, and $ 63 $ negative links with $ 4.269 $ weight) and the after period ($ 206 $ positive links with $ 21.665 $ weight, and $ 82 $ negative links with $ 6.516 $ weight). Similarly, the total weights of positive ($ 23.005 $) and negative ($ 5.482 $) subnetworks in the European debt crisis are notably higher than those weights after (positive $ 22.057 $ and negative $ 3.762 $) the crisis.} Arguably, the more significant connection and resonance effect are partly derived from the `spillover effect' that national equity markets develop more cross-continent relations beyond the information flow.\footnote{In the subprime crisis, $ 11 $ positive links with total weights of $ 0.935 $ flow from Asia to Americas ($ 10 $ edges with $ 0.513 $ before the crisis and six edges with $ 0.505 $ after the crisis), 30 positive links with 2.214 flow from Europe to Asia (13 edges with $ 0.775 $ before the crisis and $ 23 $ edges with $ 1.607 $ after the crisis), and $ 20 $ positive links with $ 3.439 $ flow from Americas to Europe ($ 26 $ edges with $ 1.692 $ before the crisis and $ 15 $ edges with $ 0.944 $ after the crisis). In the European debt crisis, $ 16 $ positive links with total weights of $ 0.942 $ flow from Asia to the Americas ($ 15 $ edges with $ 0.606 $ after the crisis), and $ 29 $ positive links with total weights of $ 1.777 $ flow from Europe to Asia ($ 33 $ edges with $ 1.627 $ after the crisis). As the only exception, more positive links with higher weights flow from the Americas to Europe after the European debt crisis because the United States market restarts to interact with European markets in this period. (See Figure \ref{Fig HeatMapVARCoefs_EuDebtAfter} for details.)} The spillover effect is also significant in the negative subnetwork during two crises, revealed in more negative connections between continents.\footnote{Cross-continent links' total numbers and weights (excluding links in the information flow) reach the maximum values at $ 45 $ and $ 3.132 $ in the subprime crisis, respectively. In contrast, only one link with a weight of $ 0.133 $ exists before the crisis, and $ 13 $ links with $ 0.767 $ after the crisis. Moreover, the total number and weight of cross-continent links are $ 22 $ and $ 1.879 $ during the European debt crisis, notably higher than the total link number of $ 17 $ and weight of $ 0.735 $ after the crisis.} The spillover effect mirrors an unusual feature of the global equity system in the crisis that not only national markets develop closer international relations along with the information flow but also build cross-continent connections beyond the flow.

Our analysis are consistent with three strands of literature. The first strand of literature concerns the demand shocks during the crisis \citep{forbes2002no, forbes2004asian}. The negative demand shocks decrease the stock prices of all national equity markets and result in co-movements with the same direction, affecting countries that export to regions with a more sophisticated economy and multiple supply chains. The second strand of literature concerns the role of cash-flow shock during the crisis \citep{bekaert2014global, boyer2006how}. The productivity shock experienced in a country should increase the co-movement of its stock prices in the same supply chain but decrease the co-movement with those in competing supply chains. The final strand of literature concerns the currency channel of exchange and changes in trading following the crisis \citep[see][for detailed explanations]{albuquerque2009global, hau2017the}. For instance, the negative equity shock experienced in Japan during the crisis led to a relative depreciation in Yen, as implied by \citet{pavlova2007asset}. Japan would become more competitive compared with its competitor China. All else equal, Chinese and Japanese stock prices should, in theory, move in the opposite direction, as we have shown in the interactions in the negative subnetwork.

%===================================================================================%
%===================================================================================%
%===================================================================================%
%===================================================================================%
%===================================================================================%

\subsection{National Equity Markets Analysis}
\label{Section Country Status Analysis Based on Strengths}

The net in- and out-strengths defined in Equation~\eqref{Eq Net In- and Out- Strenghts} are investigated to demonstrate how a given national stock market interacts with the global equity market. Positive (negative) values of net in-strengths reflect homogenous (heterogeneous) effects from the global system, while national equity markets respond to such impulses through positive (negative) out-strengths. For a given national market $ k $, there are four possible interactive patterns in the global market: $ \mathrm{NetOutS}(k) > 0 $ and $ \mathrm{NetInS}(k) > 0 $, $ \mathrm{NetOutS}(k) < 0 $ and $ \mathrm{NetInS}(k) > 0 $, $ \mathrm{NetOutS}(k) < 0 $ and $ \mathrm{NetInS}(k) < 0 $, and $ \mathrm{NetOutS}(k) > 0 $ and $ \mathrm{NetInS}(k) < 0 $, corresponding to four quadrants in Figures \ref{Fig Net In-Strengths and Out-Strengths Subprime} and \ref{Fig Net In-Strengths and Out-Strengths EuDebt}.

Figures \ref{Fig Net In-Strengths and Out-Strengths Subprime} and \ref{Fig Net In-Strengths and Out-Strengths EuDebt} show that nearly all listed national equity markets in Table \ref{Tab National Indexes} are located in either the first quadrant or the second quadrant over all five periods. Economic globalization is the root cause of this phenomenon: active international business drives countries to integrate into the international economic system such that the global economy has a homogenous effect on them, and different markets tend to fluctuate in the same direction. Expansions (recessions) of the global market increase (decrease) national stock markets' indexes to some degree, mirrored by positive net in-strengths. Different from net in-strengths, net out-strengths measure the extent of influence from countries on the global market. Positive out-strengths act as channels to transmit information in normal periods but potentially intensify the resonance effect in the crisis. There is one exception: China. Due to the relatively isolated stock market from the world, China is the only country with zero net in-strengths before and after the subprime crisis and after the European debt crisis.

\bigskip
\centerline{\bf [Place Figures \ref{Fig Net In-Strengths and Out-Strengths Subprime} and \ref{Fig Net In-Strengths and Out-Strengths EuDebt} about here]}
\bigskip

Based on net in- and out-strengths, Figures \ref{Fig Net In-Strengths and Out-Strengths Subprime} and \ref{Fig Net In-Strengths and Out-Strengths EuDebt} present the locations of $ 36 $ national markets to further explore their respective status in the global equity market over five periods.

\noindent
\textbf{Before the subprime crisis},
\begin{itemize}
\item Quadrant 1: Australia, Malaysia, Indonesia, Korea, Japan, Singapore, New Zealand, Thailand, China, Hong Kong, Netherlands, Greece, Belgium, France, Germany, Italy, Denmark, Norway, Sweden, Portugal, Switzerland, United Kingdom, Poland,	Austria, Brazil, Chile, Argentina, Mexico, Canada, United States;
\item Quadrant 2: 
Philippines, Spain, Ireland, Russia, and Turkey;	
\end{itemize}
\textbf{During the subprime crisis},
\begin{itemize}
\item Quadrant 1: Malaysia, Indonesia, Korea, Japan, Singapore, New Zealand, Philippines, Thailand, Hong Kong, Netherlands, Greece, Belgium, Germany, Denmark, Norway, Russia, Switzerland, Poland, Brazil, Chile, Mexico, United States;	
\item Quadrant 2: Australia, China, France, Finland, Spain, Ireland, Italy, Sweden, Portugal, United Kingdom, Turkey, Austria, Argentina, Canada;
\end{itemize}
\textbf{After the subprime crisis},
\begin{itemize}
\item Quadrant 1: Malaysia, Indonesia, Singapore, New Zealand, Philippines, Thailand, Netherlands, Greece, Belgium, France, Germany, Finland, Italy, Norway, Sweden, Poland, Turkey, Brazil, Chile, Argentina, Mexico, Canada, United States;	
\item Quadrant 2: Australia, Korea, Japan, China, Hong Kong, Spain, Ireland, Denmark, Portugal, Russia, Switzerland, United Kingdom, Austria;
\end{itemize}\textbf{During the European debt crisis},
\begin{itemize}
\item Quadrant 1: Australia, Malaysia, Indonesia, Korea, Japan, Singapore, Philippines, Thailand, China, Hong Kong, France, Spain, Ireland, Italy, Norway, Sweden, Portugal, Russia, Switzerland, United Kingdom, Poland, Turkey, Austria, Brazil, Chile, Argentina, Canada, United States;	
\item Quadrant 2: New Zealand, Netherlands, Greece, Belgium, Germany, Finland, Denmark, Mexico;
\end{itemize}\textbf{After the European debt crisis},
\begin{itemize}
\item Quadrant 1: Australia, Malaysia, Indonesia, Korea, Japan, Singapore, Philippines, Thailand, China, Hong Kong, Netherlands, France, Germany, Finland, Spain, Italy, Norway, Portugal, Russia, United Kingdom, Turkey, Austria, Brazil, Chile, Argentina, Mexico, Canada, United States;	
\item Quadrant 2: New Zealand, Greece, Belgium, Ireland, Denmark, Sweden, Switzerland, Poland.
\end{itemize}

Eight national equity markets are in the first quadrant throughout the five periods, including Malaysia, Indonesia, Singapore, Thailand, Norway, Brazil, Chile, and United States. The stable statuses in the first quadrant show that stock markets in these countries are highly consistent with the global market over the entire period. It is not surprising that the United States market stays in the first quadrant because of its leading role in the global financial system and its predominant influence on the world economy. Moreover, the Singapore market is one of the few with the top net out-strengths in the entire period, acting as a critical intermediary to transmit information in the global market. Top net out-strengths are partly due to Singapore's irreplaceable role in the global capital market and its unique developed country status in Southeast Asia. Top net out-strengths are also partly due to the relatively late closing time of the Singapore market, making it receive more information from other Asian markets like Japan, Korea, and Hong Kong. For comparison, the global market poses more significant effects on developing countries like Malaysia, Indonesia, Brazil, and Chile, forcing fluctuations in these equity markets to follow global trends.

However, the subprime and European debt crises give rise to status changes in many national markets. Notably, Korea, Japan, and Hong Kong are located in the second quadrant only after the subprime crisis and stayed in the first quadrant over the rest four periods. The above quadrant changes are because these markets are no longer part of the information flow and have little influence on the following continents. Moreover, Netherlands, France, Germany, Italy, Argentina, Mexico, and Canada temporarily move to the second quadrant during the subprime crisis or the European debt crisis, causing negative feedback within continents. Compared with positive net out-strengths, the negative ones are beneficial for stabilizing the global system in the crisis period since they can alleviate the resonance effect by preventing national equity markets from homogeneous behaviors (i.e., the synchronous declines in national composite indexes). Analogous to France and Italy, Australia, China, Portugal, United Kingdom, and Austria appear in the second quadrant during the subprime crisis and return to the first quadrant during and after the European debt crisis. Nonetheless, the subprime crisis has a more profound influence on the latter five countries because they are still in the second quadrant after the crisis.\footnote{The strength analysis matches well with the severity of the equity market decline during the subprime and European debt crises. It shows that those countries with the strongest equity market that collapsed during the crisis have historically less exposure to the global market. Likewise, the less severely affected national equity markets generally have substantially higher loadings for several global factors before the crisis \citep[e.g.,][]{hong1999unified, hong2007industries}.}

Our empirical results are consistent with the literature on information frictions due to limited attention and limited information processing capabilities on the part of investors, causing the equity prices in certain countries to underreact to information relevant to broader economic conditions \citep[e.g.,][]{hong1999unified, hong2007industries}. As United States has the world's largest GDP and is an important trading partner for many countries, shocks to its economy have necessary knock-on effects in other industrialized markets. The United States equity market is also the world's most significant in terms of market capitalization and likely receives the most attention from investors. Thus, information obtained during the subprime crisis diffuses gradually from the United States equity market to other national markets, consistent with the Lucas-tree model with gradual cross-country information diffusion.

%===================================================================================%
%===================================================================================%
%===================================================================================%
%===================================================================================%
%===================================================================================%
%===================================================================================%
%===================================================================================%
%===================================================================================%
%===================================================================================%
%===================================================================================%

\section{Dynamic Estimation of the Global Equity Market over Long Period}
\label{Section: Dynamic Estimation}

So far, the time-zone VAR model treats the comovements among national equity markets, reflected as VAR coefficients, as stable interactions in sample periods. To further demonstrate the discovered contagion mechanism in the global equity market over a long period, we investigate the continent strengths from 2001 to 2022 in a dynamic view.

We use the rolling-window estimation, with the repeated improved CV defined in Equation~\eqref{Eq Optimal Lambdas for Adjacency Matrices} to determine tuning parameters in each window. Following \cite{demirer2018estimating}, we set a fixed window length of $ T = 150 $ days (about half of one trading year) to study the global equity network over two decades from 8 January 2001 to 18 November 2022. In each window, we move the beginning point forward by a fixed interval of $ \Delta T = 5 $ days (about one trading week) until the whole period is covered. Then, we take a snapshot in each time window and generate the corresponding network. Besides, the year label is marked at the first rolling window entering the corresponding year.\footnote{For example, the rolling window starting on 1 January 2018 and ending on 27 July 2018 is labeled as `2018'.}

\bigskip
\centerline{\bf [Place Figures \ref{Fig Dynamic Continent Strength---Positive} and \ref{Fig Dynamic Continent Strength---Negative} about here]}
\bigskip

Figures \ref{Fig Dynamic Continent Strength---Positive} and \ref{Fig Dynamic Continent Strength---Negative} confirm the contagion pattern in the global market in Section \ref{Section Static Estimation} from a dynamic view. Figure \ref{Fig Dynamic Continent Strength---Positive} presents the dynamics of the positive strengths within and between three continents over a long period and suggests that continent relations in the information flow have significantly higher positive strengths than other relations most time. As shown in Figure \ref{Fig Dynamic Continent Strength---Negative}, the dynamic negative strengths illustrate that over most of the long period, the majority of negative relations exist in the interiors of continents, and the negative feedback is the dominant pattern within each continent. Moreover, the spillover effect exists during two crisis periods. Figure \ref{Fig Dynamic Continent Strength---Positive} demonstrates the abnormal growths in the positive strengths of cross-continent relations beyond the information flow in crisis periods. Figure \ref{Fig Dynamic Continent Strength---Negative} suggests the surge of negative strengths outside each continent, implying the more significant spillover effect in the negative subnetwork during two crises.

It is worth noting that the rolling window, starting on 30 July 2007 and ending on 22 February 2008, witnesses remarkable surges in positive and negative continent strengths from Asia to Europe, Europe to Europe, and Americas to Europe. These surges are mainly due to the subprime crisis. As discussed in Section \ref{Appendix A sample divisions}, this rolling window corresponds to the first half period of the subprime crisis, and a series of landmark events in August 2007 eventually contribute to the global equity market plunges.

Figures \ref{Fig Dynamic Continent Strength---Positive} and \ref{Fig Dynamic Continent Strength---Negative} also present the changes in continent strengths during the COVID-19 crisis. Notably, the COVID-19 crisis has profoundly impacted the global equity market. Dramatic increases caused by the COVID-19 crisis can be found in positive and negative continent strengths from Asia to Asia, Asia to Europe, Europe to Asia, Europe to Europe, Europe to Americas, and Americas to Europe. These remarkable growths reveal that the COVID-19 crisis poses such significant impacts on national equity markets that the global market exhibits abnormal behaviors and opens up new channels for information transmission beyond the traditional contagion mechanism. Furthermore, the impacts of the COVID-19 crisis are more sudden and severe than those of the subprime and European debt crises, consistent with the evidence in a series of influential papers that COVID-19 acts more like a short-term shock rather than a long-term crisis \citep{caballero2021model, duchin2021covid, barry2022corporate,huang2020effective,guerrieri2022macroeconomic, eichenbaum2021macroeconomics, acemoglu2021optimal, baqaee2022supply, augustin2022sickness}.

Despite the potential of the time-zone VAR model to investigate the global market during the COVID-19 crisis, the exogenous effect of the pandemic has discouraged its further utilization. Unlike the subprime and European debt crises, the COVID-19 crisis is derived from an external shock, causing massive panic in stock markets and prompting governments to announce a series of economic policies. Therefore, it is essential to consider exogenous variables related to daily COVID-19 infections (e.g., daily new cases or daily deaths) for in-depth analyses of the crisis. However,  such exogenous variables are typically non-stationary time series, and hence, they cannot be directly incorporated into equations under the VAR specification.
% However, COVID-19's exogenous effect discourages us from further utilizing the time-zone VAR model to investigate the global market during the COVID-19 crisis. Unlike the subprime and European debt crisis, the COVID-19 crisis derives from the external shock--the COVID-19 pandemic--on the financial system that causes massive panic in stock markets and forces governments to announce a series of economic policies eventually stimulating national equity markets. Therefore, considering exogenous variables on COVID-19 (e.g., daily new cases or daily deaths) is vital to in-depth analyses of the COVID-19 crisis. Nevertheless, exogenous variables are typically non-stationary time series and cannot be directly incorporated into equations under the VAR specification.

Due to the drawbacks of the rolling window method, we emphasize static estimation results more than dynamic ones. Notably, the rolling window method equally treats samples in each period, thus only describing sample features in a time interval but failing to characterize the information at a specific time. Therefore, it is reasonable to use the rolling window method to investigate the evolution trend of the global market. Still, performing local analysis in short periods may not be reliable, especially for in-depth analyses of the COVID-19 shock to the global stock market. Consequently, we recommend using more sophisticated econometric methods, such as regression discontinuity design or localized nonparametric models, to study how equity markets change in short periods.

\section{Additional Results}
\label{Section: Additional Results}

In this section, we demonstrate that taking time-zone effect and comovement directions into account can reveal more information on the global equity market by comparing the estimation results of classic and time-zone models and by comparing the network properties of unsigned networks with signed ones.

\subsection{Comparison between Classic and Time-Zone VAR Models}
\label{Section Comparison between Time-Zone and Classic VAR Models}

We compare the proposed time-zone VAR model~\eqref{Eq Time-Zone VAR Model} with the classic model, defined as
\begin{align}
\left\lbrace
\begin{array}{cll}
	r_{i_0 t} = &\sum\limits_{i \,\in\, As} \beta_{i_0 i} r_{i, t-1} + \sum\limits_{j \,\in\, Eu} \beta_{i_0 j} r_{j, t-1} + \sum\limits_{k \,\in\, Am} \beta_{i_0 k} r_{k, t-1} + \varepsilon_{i_0 t},  &  i_0 \in As, \\
	r_{j_0 t} = &\underline{\sum\limits_{i \,\in\, As} \beta_{j_0 i} r_{i, t-1}} + \sum\limits_{j \,\in\, Eu} \beta_{j_0 j} r_{j, t-1} + \sum\limits_{k \,\in\, Am} \beta_{j_0 k} r_{k, t-1} + \varepsilon_{j_0 t},  &  j_0 \in Eu, \\
	r_{k_0 t} = &\underline{\sum\limits_{i \,\in\, As} \beta_{k_0 i} r_{i, t-1} + \sum\limits_{j \,\in\, Eu} \beta_{k_0 j} r_{j, t-1}} + \sum\limits_{k \,\in\, Am} \beta_{k_0 k} r_{k, t-1} + \varepsilon_{k_0 t},  &  k_0 \in Am,
\end{array}
\right.
\label{Eq Classic VAR Model}
\end{align}
to illustrate the advantages of our model in describing the daily global equity market. The differences between the time-zone and classic VAR model derive from the latest information on Asian and European markets, underlined in Equation~\eqref{Eq Classic VAR Model}.

For a given national equity market $ l $, the in-sample ratio is used to evaluate the in-sample performance, defined as
\begin{align}
R_{IS, \, l}^2 = 1 - \frac{ \sum_{t=2}^{T} ( r_{lt} - \widehat{r}_{lt}^{\mathrm{tz}} )^2 }{ \sum_{t=2}^{T} ( r_{lt} - \widehat{r}_{lt}^{\mathrm{cls}} )^2 },
\label{Eq R2 Insample Performance}
\end{align}
where $ \widehat{r}_{lt}^{\mathrm{tz}} $ and $ \widehat{r}_{lt}^{\mathrm{cls}} $ are corresponding predicted values based on the time-zone and classic VAR model, respectively. The positive value of $ R_{IS, \, l}^2 $ means the time-zone model performs better than the classic one in depicting nation $ l $. Moreover, the larger the $ R_{IS, \, l}^2 $ is, the better the time-zone model performs.\footnote{Our results are robust when we use the out-of-sample criterion to evaluate the performance between time-zone and classic VAR models. Due to space limitations, the relevant results are available upon request.}

\bigskip
\centerline{\bf [Place Figures \ref{Fig R2 In-Sample Subprime} and \ref{Fig R2 In-Sample EuDebt} about here]}
\bigskip

As shown in Figures \ref{Fig R2 In-Sample Subprime} and \ref{Fig R2 In-Sample EuDebt}, the in-sample ratios of Asian markets over five periods are very close to zero due to the same specification on Asia in Equations~\eqref{Eq Time-Zone VAR Model} and \eqref{Eq Classic VAR Model}. In the time-zone and classic VAR model, Asian stock markets are the earliest to trade and only receive the information from the last day. The time-zone effect has a trivial impact on Asian markets and hence, fails to improve the in-sample performance of these markets. By contrast, the ratios of European and American markets are higher than zeros in five periods, showing the significantly better in-sample performance of the time-zone VAR model than the classic model. This is because the time-zone effect is notable in Europe and the Americas, such that involving the latest information on the global equity market can considerably increase the interpretability of the VAR model.

More importantly,  the average ratio of the Americas ($ 0.447 $) is about twice as much as that of Europe ($ 0.229 $) because of the more significant time-zone effect in American equity markets. Given that American markets can also receive the latest European information, higher increases in in-sample ratios can be found in American markets than in European ones, implying the better in-sample performance of American markets under the time-zone VAR specification.\footnote{Our findings are also robust to all periods, demonstrating the superiority of the time-zone VAR model over the classic one.} Therefore, the time-zone effect is vital to understanding the transmission mechanism of the daily global stock market, and ignoring this effect may lead to a considerable loss of information and make specious conclusions.

%===================================================================================%
%===================================================================================%
%===================================================================================%
%===================================================================================%
%===================================================================================%

\subsection{Comparison between Unsigned and Signed Networks}
\label{Section Comparison between Unsigned and Signed Networks}

It is worth noting that taking link signs that characterize comovement directions into account can help scholars better understand the global equity market's structural changes in different periods by providing more detailed information. To demonstrate this, we compare the topological properties of unsigned networks with those of signed ones.

First, Figure \ref{Fig Densities Before, During, After the Crisis} suggests that the unsigned density experiences substantial growth during the subprime crisis. However, such changes can be interpreted differently; stock prices may experience massive declines during the crisis \citep{calomiris2012stock} or the active self-regulatory mechanism in some national equity markets against the crisis \citep{almeida2012corporate}. Densities in signed subnetworks suggest massive declines, and self-regulatory mechanisms simultaneously lead to this sudden rise because of the notable increases in positive and negative subnetworks' densities. Moreover, the self-regulatory mechanism is more responsive to the subprime crisis because the negative network's density has a higher rate of increase in this period.

Second, we compare unsigned assortativities with signed ones in Table \ref{Tab Assortativity for Continents and Degrees}. The negative continent assortativities of the unsigned network $ \bm{A}^U $ demonstrate that national equity markets tend to develop relations with those from different continents. However, Section \ref{Section Mixing Patterns} points out that such cross-continent connections mainly appear in the positive subnetwork $ \bm{A}^+ $ to transmit information. More importantly,  the negative subnetwork $ \bm{A}^- $ has a significantly different structure from $ \bm{A}^+ $, reflected as the self-regulatory mechanism within continents. Although degree assortativities are all negative in the unsigned network, positive and negative subnetworks, two signed subnetworks can provide more information. In the subnetwork $ \bm{A}^+ $, high-degree national equity markets tend to connect low-degree markets from other continents, accelerating the massive spread of financial risks. In the subnetwork $ \bm{A}^- $, the same topological structure mainly exists at the continent level, meaning that the self-regulatory mechanism may fail to protect the system from the large-scale resonance effect. Consequently, calculating continent and degree assortativities without telling link signs may lead to specious conclusions.

Third, Tables \ref{Tab Continent Degrees Analysis} and \ref{Tab Continent Strength Analysis} demonstrate that the information flow is insignificant in the unsigned network. This insignificance is due to the interference from negative relations within continents over all periods and the temporal spillover effect triggered by the crisis. Hence, scholars may mix positive and negative subnetworks' features without distinguishing comovement directions and eventually fail to identify those critical patterns.

%===================================================================================%
%===================================================================================%
%===================================================================================%
%===================================================================================%
%===================================================================================%
%===================================================================================%
%===================================================================================%
%===================================================================================%
%===================================================================================%
%===================================================================================%

\section{Conclusions}
\label{section: conclusion}

%Global financial integration does produce superior capital allocations but also amplifies systemic risk to some degree. Yet, this requires--from economists, policy-makers, and practitioners alike--understanding of comovements occurring in the system. 

In this paper, we investigate the comovements between national equity markets from the network perspective. We introduce the time-zone effect in the VAR model to take into account the influence caused by different trading windows of national stock markets. Using data from 36 daily national stock indexes, we apply the proposed model to study how stock markets in different countries interact with others for the subprime, European debt, and COVID-19 crises. We investigate the directions of comovements revealed in signs of VAR coefficients and discover the resonance effect in the global system. During regular periods, the resonance effect is reflected as the information flow occurring in the global system that fluctuates in national equity markets -- beginning in Asia, moving to Europe, and finishing in the Americas. By contrast, the resonance effect is more significant during the subprime,  European debt, and COVID-19 crises, as reflected in excessive comovements with the same direction beyond the information flow. Our results on densities and assortative coefficients suggest abnormal changes in the transmission mechanism of the global system during these crises. Strength analysis further demonstrates the distinct roles that different national stock markets played in the global market over crisis periods. More importantly, compared with unsigned networks, signed networks can provide more information on the global equity market and better identify those critical contagion patterns. Overall, our work identifies the global system transmission channels of financial contagion.

%===================================================================================%
%===================================================================================%
%===================================================================================%
%===================================================================================%
%===================================================================================%

\clearpage

{
\small
\singlespacing
\bibliographystyle{apalike}
\bibliography{ReferencesTZVARModelsLASSO}
}

%%% FIGURE 1 %%%
\clearpage

%%% FIGURE 2 %%%
\clearpage
\begin{figure}
\begin{center}
		\includegraphics{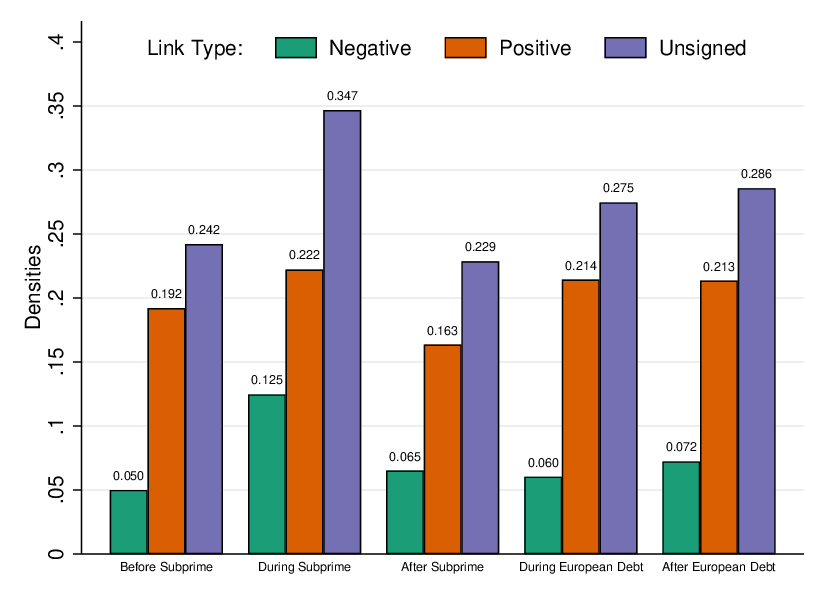}
	\caption{\textbf{Network densities in five periods.} This figure shows network densities before, during, after the subprime crisis, during, and after the European debt crisis. Purple, orange, and green bars reflect the densities of the unsigned network, and positive and negative subnetworks, respectively, and the numbers above bars show the corresponding density values. Unsigned, positive and negative densities are defined in Equations \eqref{Eq Density of Unsigned Network} to \eqref{Eq Density of Negative Subnetwork}.}
	
	\label{Fig Densities Before, During, After the Crisis}
\end{center}
\end{figure}

%%% FIGURE 3 %%%
\clearpage
\begin{figure} 
\begin{center}
	\subfloat{\includegraphics[scale=0.65]{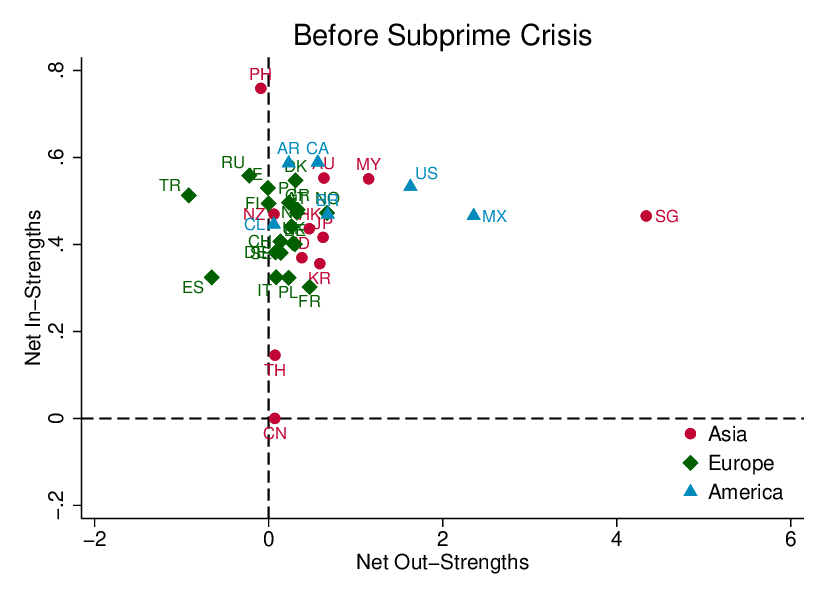}} \\
	\subfloat{\includegraphics[scale=0.65]{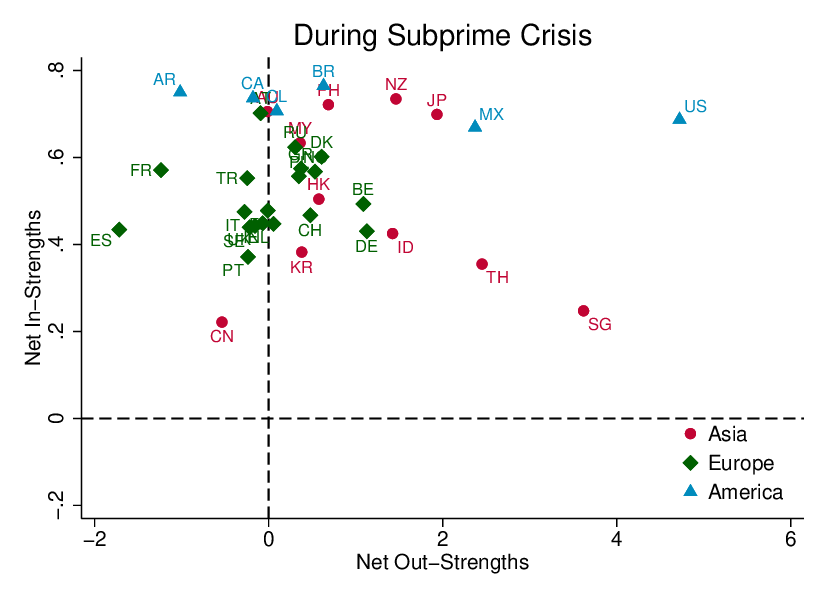}} \\
	\subfloat{\includegraphics[scale=0.65]{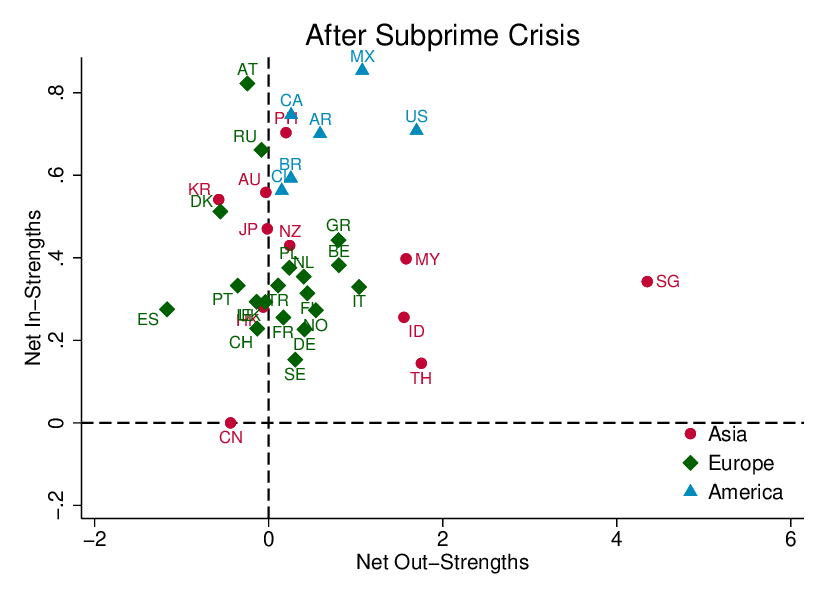}}
	
	\caption{\textbf{Net in-strengths and out-strengths over three periods in the subprime crisis.} In each scatter plot, the $ x $-axis is the net out-strength and the $ y $-axis is the net in-strength. Red circle, green diamond, and blue triangle represent equity markets from Asia, Europe and Americas. Net in- and out-strengths are defined in Equation \eqref{Eq Net In- and Out- Strenghts}.}
	
	\label{Fig Net In-Strengths and Out-Strengths Subprime}
\end{center}
\end{figure}

%%% FIGURE 4 %%%
\clearpage
\begin{figure} 
\begin{center}
	\subfloat{\includegraphics[scale=0.65]{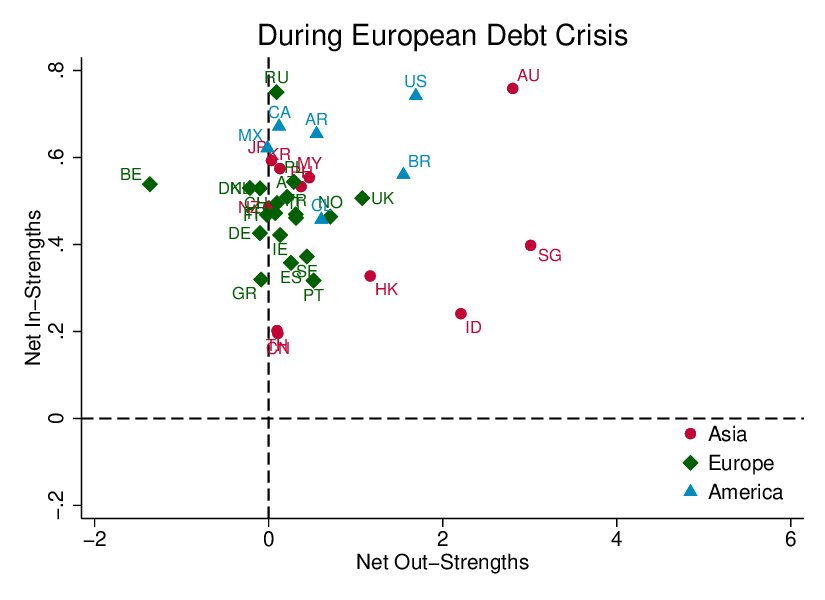}} \\
	\subfloat{\includegraphics[scale=0.65]{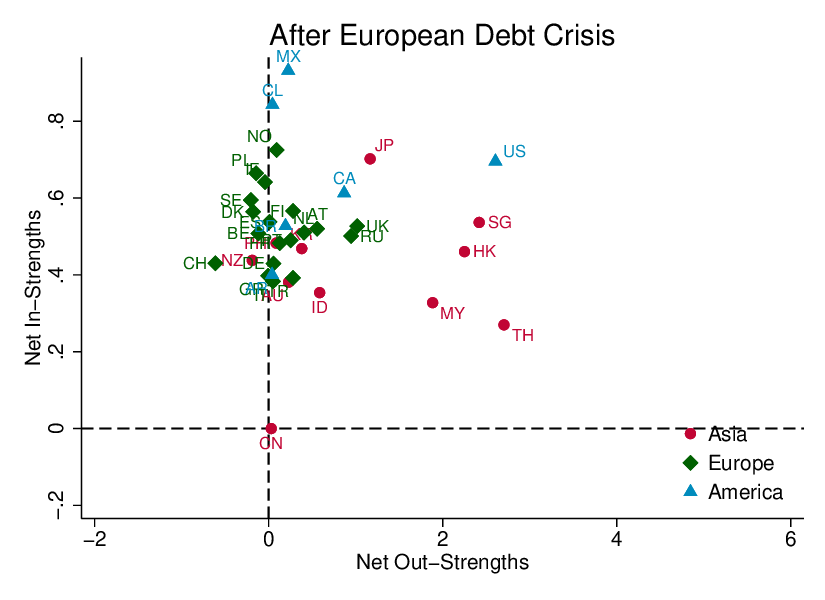}}
	
	\caption{\textbf{Net in-strengths and out-strengths over two periods in the European debt crisis.} In each scatter plot, the $ x $-axis is the net out-strength and the $ y $-axis is the net in-strength. Red circle, green diamond, and blue triangle represent equity markets from Asia, Europe and Americas. Net in- and out-strengths are defined in Equation \eqref{Eq Net In- and Out- Strenghts}.}
	
	\label{Fig Net In-Strengths and Out-Strengths EuDebt}
\end{center}
\end{figure}

%%% FIGURE 5 %%%
\clearpage
\begin{landscape}
\begin{figure}
\begin{center}
	\makebox[\textwidth]{\includegraphics[scale=0.35]{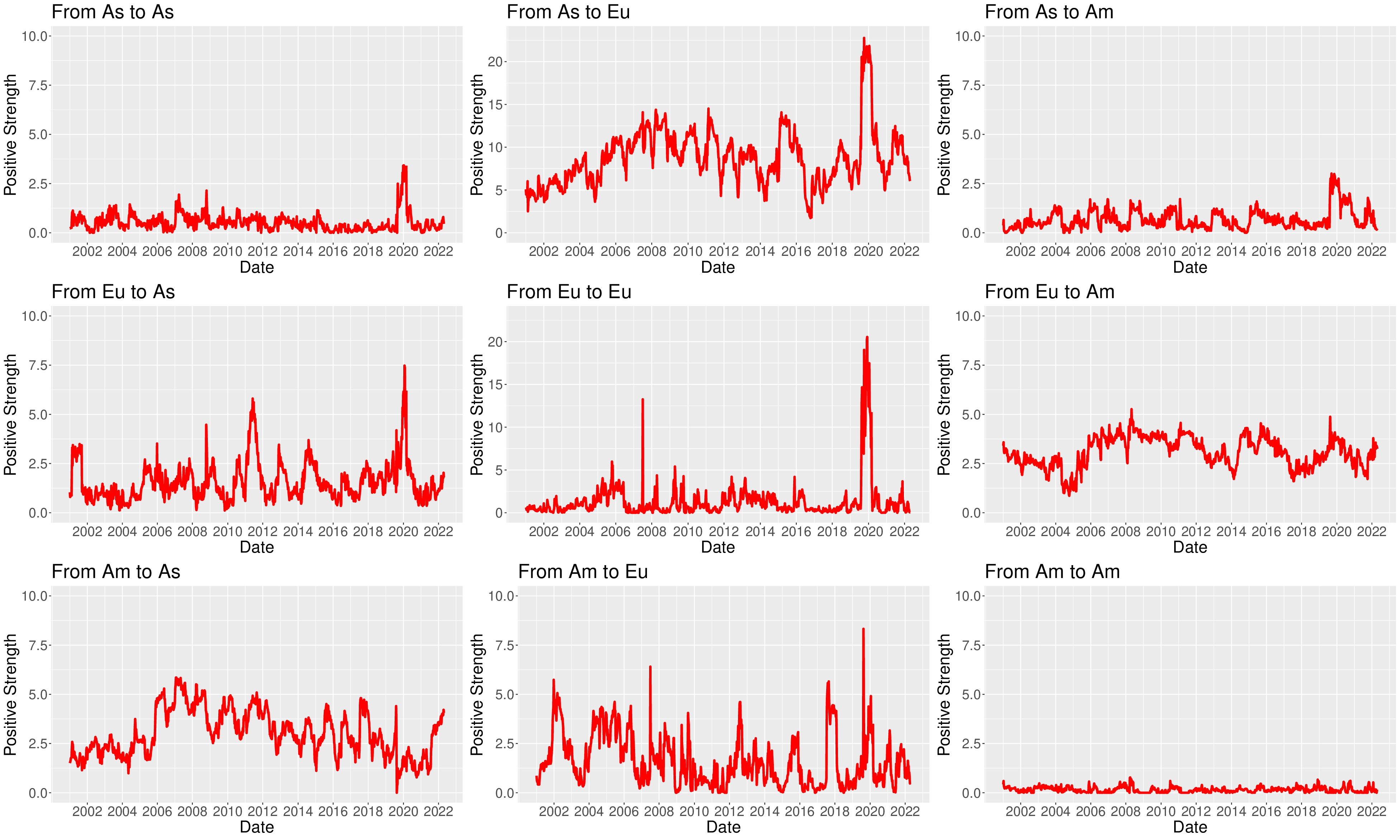}}
	
	\caption{\textbf{The dynamics of continents' positive strengths in the global equity market.} The year label is marked at the first rolling window entering the corresponding year, and positive continent strengths are defined in Equation \eqref{Eq Continent Strength Positive}.}
	
	\label{Fig Dynamic Continent Strength---Positive}
\end{center}	
\end{figure}
\end{landscape}

%%% FIGURE 6 %%%
\clearpage
\begin{landscape}
\begin{figure}
\begin{center}
	\makebox[\textwidth]{\includegraphics[scale=0.35]{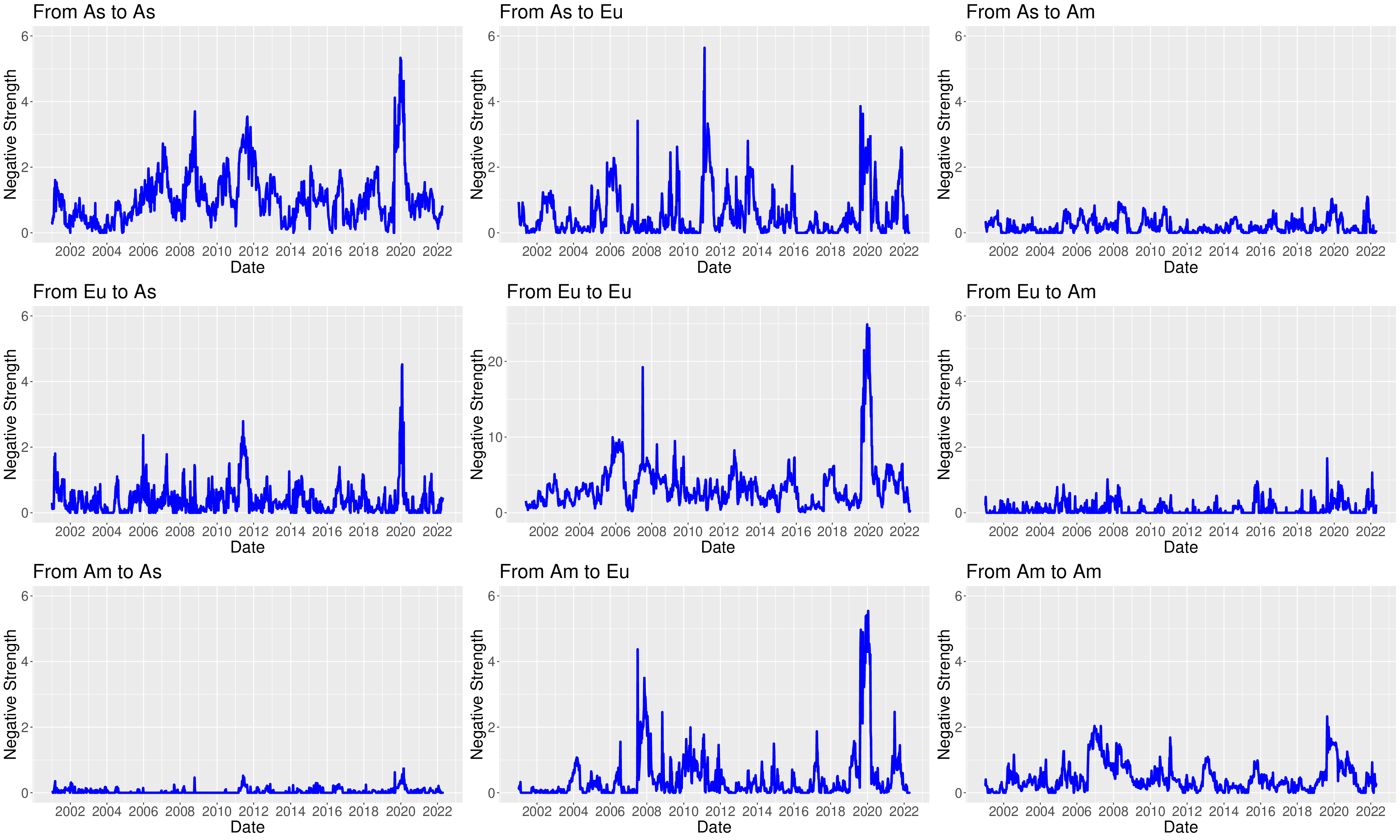}}
	
	\caption{\textbf{The dynamics of continents' negative strengths in the global equity market.} The year label is marked at the first rolling window entering the corresponding year, and negative continent strengths are defined in Equation \eqref{Eq Continent Strength Negative}.}
	
	\label{Fig Dynamic Continent Strength---Negative}
\end{center}	
\end{figure}
\end{landscape}

%%% FIGURE 7 %%%
\clearpage
\begin{figure}
\begin{center}
	\subfloat{\includegraphics[scale=0.69]{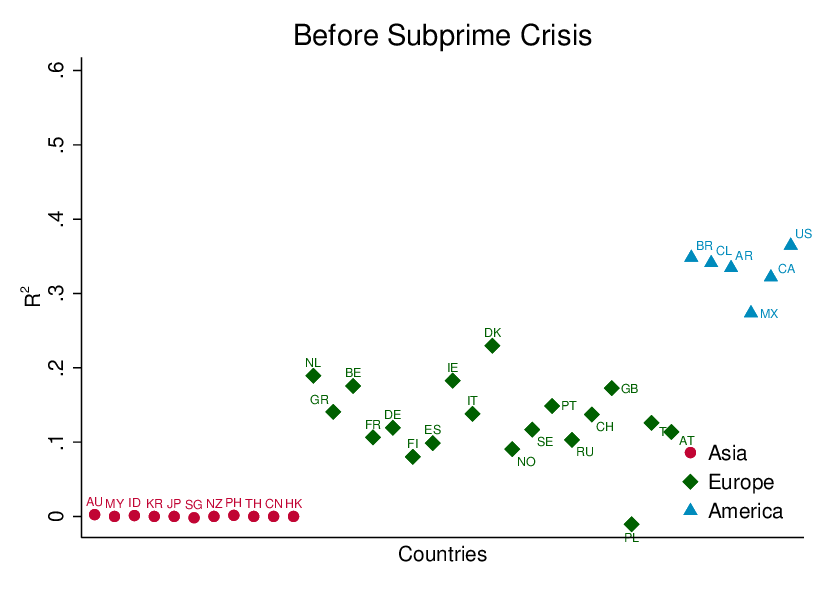}} \\
	\subfloat{\includegraphics[scale=0.69]{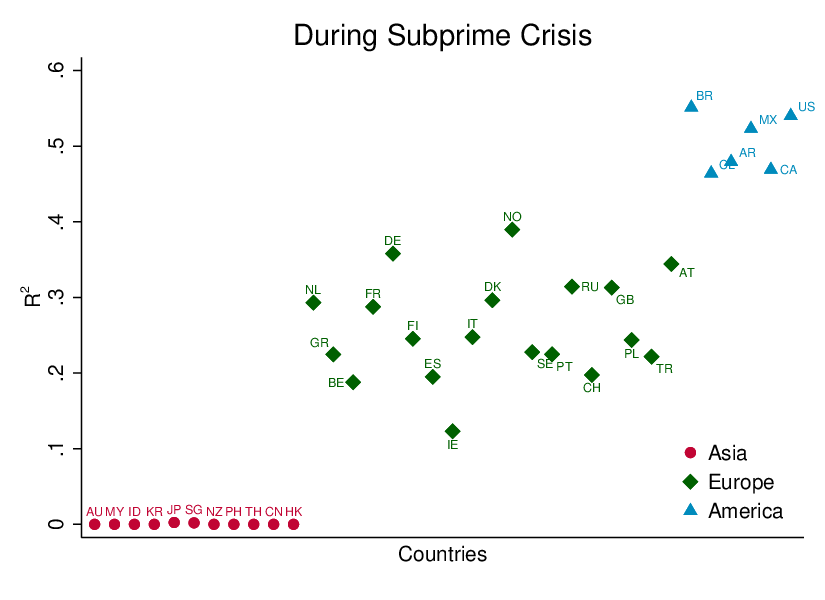}} \\
	\subfloat{\includegraphics[scale=0.69]{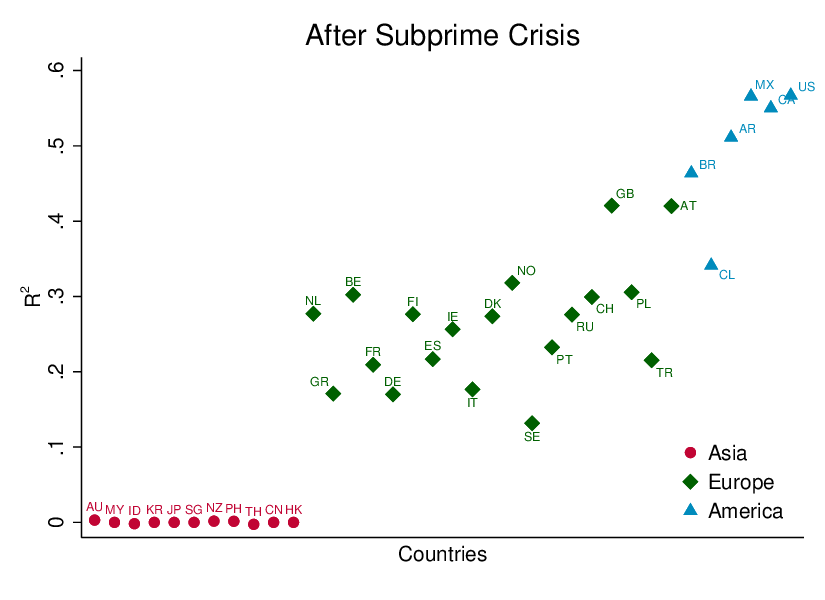}}
	
	\caption{\textbf{In-sample ratios $ R_{IS}^2 $ of national equity markets over three periods in the subprime crisis.} Red circle, green diamond, and blue triangle represent equity markets from Asia, Europe and Americas. The in-sample ratio is defined in Equation \eqref{Eq R2 Insample Performance}.}
	
	\label{Fig R2 In-Sample Subprime}
\end{center}		
\end{figure}

%%% FIGURE 8 %%%
\clearpage
\begin{figure}
\begin{center}
	\subfloat{\includegraphics[scale=0.7]{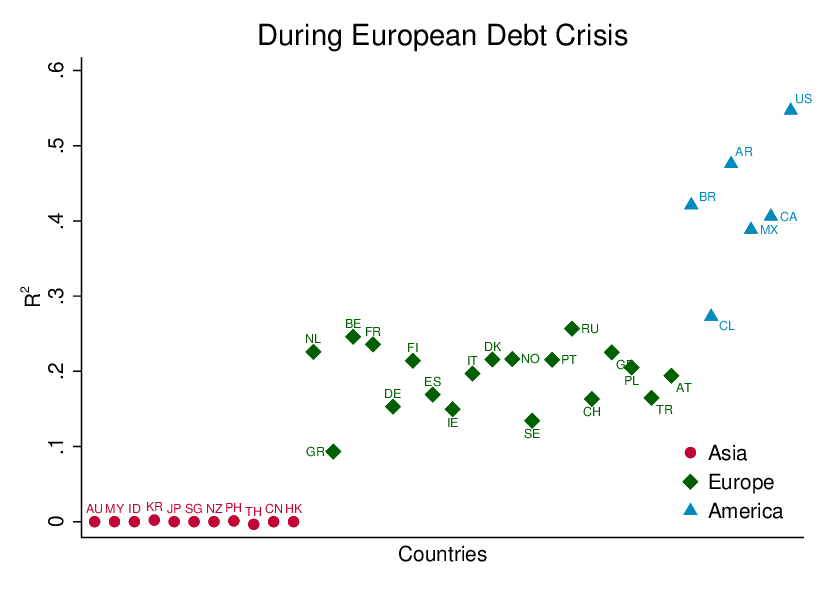}}   \\
	\subfloat{\includegraphics[scale=0.7]{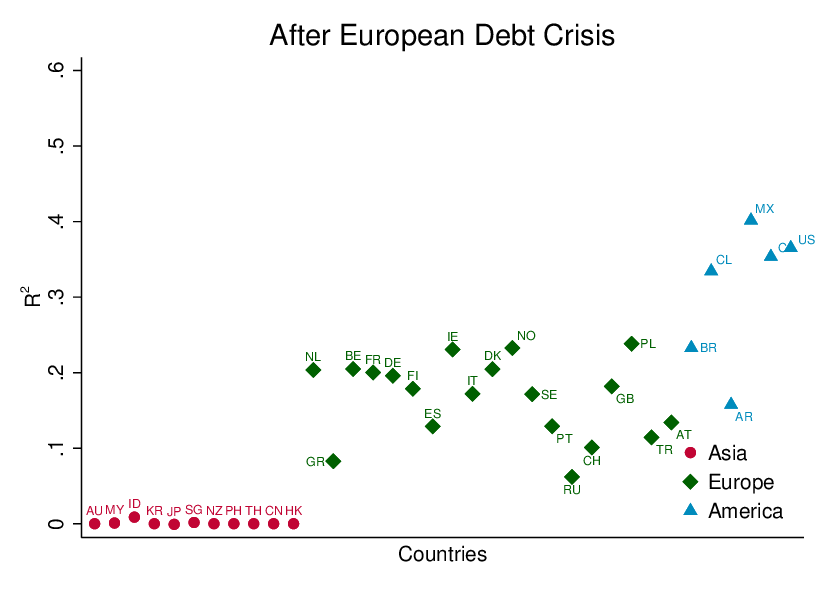}}
	
	\caption{\textbf{In-sample ratios $ R_{IS}^2 $ of national equity markets over two periods in the European debt crisis.} Red circle, green diamond, and blue triangle represent equity markets from Asia, Europe and Americas. The in-sample ratio is defined in Equation \eqref{Eq R2 Insample Performance}.}
	
	\label{Fig R2 In-Sample EuDebt}
\end{center}		
\end{figure}

%%% TABLE 1 %%%
\begin{table}
\begin{center}

\caption{\textbf{Information on national stock return indexes and equity market opening/closing times.} This table shows national stock return indexes. Column (2) provides the names of the stock return indexes; Column (3) is the opening and closing times (Eastern Standard Time; EST) for national equity markets; Column (4) indicates the continent to which the country belongs. EST: Eastern Standard Time.}

\label{Tab National Indexes}

\begin{spacing}{1}
\begin{tabular}{lccc}
	\toprule
	Country        & Index name                  & Opening/closing times (EST) & Continent \\ \hline
	Australia      & ASX 51                      & 7.00p/1.00a                 & Asia      \\
	Malaysia       & FBMKCI                      & 8.00p/4.00a                 & Asia      \\
	Indonesia      & JCI                         & 9.00p/4.00a                 & Asia      \\
	Korea          & KOSPI                       & 7.00p/1.30a                 & Asia      \\
	Japan          & NKY                         & 7.00p/1.00a                 & Asia      \\
	Singapore      & STI                         & 8.00p/4.00a                 & Asia      \\
	New Zealand    & NESE 10                     & 5.00p/0.00a                 & Asia      \\
	Philippines    & PCOMP                       & 8.30p/2.30a                 & Asia      \\
	Thailand       & SET                         & 10.00p/4.30a                & Asia      \\
	China          & SHCOMP                      & 8.30p/2.00a                 & Asia      \\
	Hong Kong      & HIS                         & 8.30p/3.00a                 & Asia      \\     \hline
	Netherlands    & AEX                         & 9.00a/11.40a                & Europe    \\
	Greece         & ASE                         & 3.00a/11.30a                & Europe    \\
	Belgium        & BEL 20                      & 3.00a/11.30a                & Europe    \\
	France         & CAC                         & 3.00a/11.30a                & Europe    \\
	Germany        & DAX                         & 2.00a/2.00p                 & Europe    \\
	Finland        & HEX                         & 3.00a/11.30a                & Europe    \\
	Spain          & IBEX                        & 3.00a/11.30a                & Europe    \\
	Ireland        & ISEQ                        & 3.00a/11.30a                & Europe    \\
	Italy          & IT 30                       & 3.00a/11.35a                & Europe    \\
	Denmark        & KFX                         & 3.00a/11.00a                & Europe    \\
	Norway         & OBXP                        & 3.00a/10.30a                & Europe    \\
	Sweden         & OMX                         & 3.00a/11.30a                & Europe    \\
	Portugal       & PSI 20                      & 3.00a/11.30a                & Europe    \\
	Russia         & RTSI                        & 2.00a/10.45a                & Europe    \\
	Switzerland    & SMI                         & 3.00a/11.20a                & Europe    \\
	United Kingdom & UKX 100                     & 3.00a/11.30a                & Europe    \\
	Poland         & WIG                         & 3.00a/11.00a                & Europe    \\
	Turkey         & XU 100                      & 2.00a/10.00a                & Europe    \\
	Austria        & ATX                         & 3.00a/11.30a                & Europe    \\     \hline
	Brazil         & IBOV                        & 8.00a/4.00p                 & America   \\
	Chile          & IPSA                        & 9.30a/4.00p                 & America   \\
	Argentina      & MERVAL                      & 9.00a/4.00p                 & America   \\
	Mexico         & MEXBOL                      & 9.30a/4.00p                 & America   \\			
	Canada         & SPTSX                       & 9.30a/4.00p                 & America   \\
	United States  & SPX                         & 9.30a/4.00p                 & America   \\    \bottomrule
\end{tabular}
\end{spacing}

\end{center}
\end{table}

%%% TABLE 2 %%%
\begin{table}
\begin{center}
\begin{spacing}{1.5}

\caption{\textbf{The continent and degree assortativities for unsigned networks, positive and negative subnetworks over five periods.} Positive values are marked in bold, and continent and degree assortativities are defined in Equations \eqref{Eq Assortatative for Continents} and \eqref{Eq Assortative for Degrees}, respectively.}
\label{Tab Assortativity for Continents and Degrees}

\begin{tabular}{lcccccc}
	\toprule
	&          & \multicolumn{3}{c}{Subprime Crisis} & \multicolumn{2}{c}{\small European Debt Crisis} \\ \cmidrule(lr){3-5}  \cmidrule(lr){6-7} 
	&          & Before     & During     & After     & During            & After             \\ \cmidrule(lr){2-7} 
	& Unsigned & -0.152     & -0.114     & -0.157    & -0.231            & -0.146            \\
	Continent  & Positive & -0.374     & -0.277     & -0.310    & -0.355            & -0.271            \\
	& Negative & \textbf{0.931}      & \textbf{0.335}      & \textbf{0.705}     & \textbf{0.339}             & \textbf{0.618}            \\ \cmidrule(lr){2-7} 
	& Unsigned & -0.204     & -0.015     & -0.189    & -0.377            & -0.119            \\
	Degree     & Positive & -0.278     & -0.086     & -0.267    & -0.438            & -0.340            \\
	& Negative & -0.279     & -0.127     & -0.164    & -0.186            & -0.188            \\ \bottomrule
\end{tabular}

\end{spacing}
\end{center}
\end{table}

%%% TABLE 3 %%%
\begin{landscape}
\begin{table}
\begin{center}
\begin{spacing}{1.5}

\caption{\textbf{Continent degrees based on adjacency matrixes}. Values in the information flow are marked in bold for the unsigned network and the positive subnetwork. Values within continents (diagonal elements) are marked in bold for the negative subnetwork. Continent degrees are computed by replacing $ w_{ij}^+ $ and $ w_{ij}^- $ in Equations \eqref{Eq Continent Strength Unsigned} to \eqref{Eq Continent Strength Negative} with $ a_{ij}^+ $ and $ a_{ij}^- $, respectively.}
\label{Tab Continent Degrees Analysis}

\begin{tabular}{lllccccccccc}
	\toprule
	&        &         &      & Unsigned &         &      & Positive &         &      & Negative &         \\ \cmidrule(lr){4-6} \cmidrule(lr){7-9}  \cmidrule(lr){10-12} 
	& Period & From\textbackslash{}to & Asia & Europe   & Americas & Asia & Europe   & Americas & Asia & Europe   & Americas \\  \cmidrule(lr){2-12} 
	&        & Asia    & 18   & \textbf{106}      & 11      & 1    & \textbf{106}      & 10      & \textbf{17}   & 0        & 1       \\
	& Before & Europe  & 13   & 35       & \textbf{43}      & 13   & 1        & \textbf{43}      & 0    & \textbf{34}       & 0       \\
	&        & Americas & \textbf{42}   & 26       & 11      & \textbf{42}   & 26       & 0       & 0    & 0        & \textbf{11}      \\ \cmidrule(lr){2-12} 
	&        & Asia    & 36   & \textbf{128}      & 19      & 9    & \textbf{127}      & 11      & \textbf{27}   & 1        & 8       \\
	Subprime Crisis & During & Europe  & 37   & 86       & \textbf{47}      & 30   & 5        & \textbf{43}      & 7    & \textbf{81}       & 4       \\
	&        & Americas & \textbf{37}   & 42       & 5       & \textbf{34}   & 20       & 1       & 3    & 22       & \textbf{4}       \\ \cmidrule(lr){2-12} 
	&        & Asia    & 20   & \textbf{89}       & 6       & 7    & \textbf{80}       & 6       & \textbf{13}   & 9        & 0       \\
	& After  & Europe  & 24   & 66       & \textbf{32}      & 23   & 13       & \textbf{32}      & 1    & \textbf{53}       & 0       \\
	&        & Americas & \textbf{30}   & 18       & 3       & \textbf{30}   & 15       & 0       & 0    & 3        & \textbf{3}       \\ \cmidrule(lr){2-12} 
	&        & Asia    & 23   & \textbf{114}      & 24      & 7    & \textbf{114}      & 16      & \textbf{16}   & 0        & 8       \\
	European Debt Crisis  & During & Europe  & 35   & 35       & \textbf{48}      & 29   & 2        & \textbf{47}      & 6    & \textbf{33}       & 1       \\
	&        & Americas & \textbf{45}   & 14       & 8       & \textbf{45}   & 7        & 3       & 0    & 7        & \textbf{5}       \\ \cmidrule(lr){2-12} 
	&        & Asia    & 21   & \textbf{115}      & 16      & 8    & \textbf{115}      & 15      & 13   & 0        & 1       \\
	& After  & Europe  & 39   & 62       & \textbf{49}      & 33   & 5        & \textbf{46}      & 6    & \textbf{57}       & 3       \\
	&        & Americas & \textbf{31}   & 21       & 6       & \textbf{31}   & 14       & 2       & 0    & 7        & \textbf{4}      \\ \bottomrule
\end{tabular}

\end{spacing}
\end{center}
\end{table}
\end{landscape}

%%% TABLE 4 %%%
\begin{landscape}
\begin{table}
\begin{center}
\begin{spacing}{1.5}

\caption{\textbf{Continent strengths based on weight matrixes}.  Values in the information flow are marked in bold for the unsigned network and the positive subnetwork. Values within continents (diagonal elements) are marked in bold for the negative subnetwork. Continent strengths are defined in Equations \eqref{Eq Continent Strength Unsigned} to \eqref{Eq Continent Strength Negative}.}
\label{Tab Continent Strength Analysis}

\begin{tabular}{lllccccccccc}
	\toprule
	&        &         &      & Unsigned &         &      & Positive &         &      & Negative &         \\ \cmidrule(lr){4-6} \cmidrule(lr){7-9}  \cmidrule(lr){10-12} 
	& Period & From\textbackslash{}to & Asia & Europe   & Americas & Asia & Europe   & Americas & Asia & Europe   & Americas \\  \cmidrule(lr){2-12} 
	&        & Asia    & 0.980 & \textbf{8.853}    & 0.646   & 0.024 & \textbf{8.853}    & 0.513   & \textbf{0.956} & 0    & 0.133   \\
	& Before & Europe  & 0.775 & 2.366    & \textbf{3.564}   & 0.775 & 0.038    & \textbf{3.564}   & 0 & \textbf{2.328}    & 0   \\
	&        & Americas & \textbf{4.679} & 1.692    & 0.852   & \textbf{4.679} & 1.692    & 0   & 0 & 0    & \textbf{0.852}   \\ \cmidrule(lr){2-12} 
	&        & Asia    & 3.107 & \textbf{12.775}   & 1.349   & 1.089 & \textbf{12.765}   & 0.935   & \textbf{2.018} & 0.011    & 0.413   \\
	Subprime Crisis & During & Europe  & 2.932 & 5.719    & \textbf{4.612}   & 2.214 & 0.415    & \textbf{4.331}   & 0.717 & \textbf{5.305}    & 0.281   \\
	&        & Americas & \textbf{5.235} & 5.063    & 0.385   & \textbf{5.149} & 3.439    & 0.064   & 0.086 & 1.624    & \textbf{0.320}   \\ \cmidrule(lr){2-12} 
	&        & Asia    & 1.437 & \textbf{9.936}    & 0.505   & 0.337 & \textbf{9.377}    & 0.505   & \textbf{1.100} & 0.559    & 0   \\
	& After  & Europe  & 1.693 & 6.202    & \textbf{3.818}   & 1.607 & 1.711    & \textbf{3.818}   & 0.086 & \textbf{4.491}    & 0   \\
	&        & Americas & \textbf{3.367} & 1.065    & 0.158   & \textbf{3.367} & 0.944    & 0   & 0 & 0.121    & \textbf{0.158}   \\ \cmidrule(lr){2-12} 
	&        & Asia    & 1.702 & \textbf{10.494}   & 1.395   & 0.556 & \textbf{10.494}   & 0.942   & \textbf{1.146} & 0    & 0.453   \\
	European Debt Crisis   & During & Europe  & 2.865 & 1.989    & \textbf{3.795}   & 1.777 & 0.115    & \textbf{3.746}   & 1.088 & \textbf{1.874}    & 0.049   \\
	&        & Americas & \textbf{4.764} & 0.796    & 0.688   & \textbf{4.764} & 0.507    & 0.105   & 0 & 0.290    & \textbf{0.583}   \\ \cmidrule(lr){2-12} 
	&        & Asia    & 1.023 & \textbf{11.155}   & 0.622   & 0.413 & \textbf{11.155}   & 0.606   & \textbf{0.610} & 0    & 0.016   \\
	& After  & Europe  & 2.043 & 2.448    & \textbf{3.821}   & 1.627 & 0.231    & \textbf{3.683}   & 0.417 & \textbf{2.216}    & 0.137   \\
	&        & Americas & \textbf{3.406} & 1.026    & 0.276   & \textbf{3.406} & 0.860    & 0.075   & 0 & 0.165    & \textbf{0.201}  \\ \bottomrule
\end{tabular}

\end{spacing}	
\end{center}	
\end{table}

\end{landscape}

%===================================================================================%
%===================================================================================%
%===================================================================================%
%===================================================================================%
%===================================================================================%

\clearpage
\appendix
\doublespacing

\begin{center}
{\bf \Large Internet Appendix}
\end{center}

% change page number and equation number in appendix
\pagenumbering{arabic}		
\renewcommand*{\thepage}{IA-\arabic{page}}

% auto change the setup for each figure/table in section 
\counterwithin*{table}{section}
\counterwithin*{figure}{section}
\setcounter{table}{0}
\setcounter{figure}{0}
\renewcommand{\thetable}{\Alph{section}.\arabic{table}}
\renewcommand{\thefigure}{\Alph{section}.\arabic{figure}}

\numberwithin{equation}{section}

\section{Methodology Details}
\label{Appendix A Methodology}

\subsection{LASSO Estimation for High-Dimensional VAR Models}

The classic VAR model fails to work in some high-dimensional cases where $ N $ is proportional to $ T $ and $ T $ goes to infinity. Arguably, the LASSO estimation is a promising solution to alleviate this problem in VAR models. We start from the classical LASSO setting. For the response variable $ \{y_t,\,t=1, \cdots, T \} $ and covariates $ \{ x_{kt},\,k=1,\cdots,p;\; t=1, \cdots, T \} $, the LASSO estimator derives from a constraint least-squares estimation \citep{tibshirani1996regression}:
\begin{align}
\widehat{\beta} = \argmin_{\beta} \sum_{t=1}^{T} \big( y_t - \sum_{k=1}^{p} \beta_k x_{kt} \big)^2 \quad s.t. \quad \sum_{k=1}^{p} \big| \beta_k \big| \leq c,
\label{Eq LASSO Estimation Constraint}
\end{align}
which is equivalent to
\begin{align}
\widehat{\beta} = \argmin_{\beta} \bigg\{ \frac{1}{2} \sum_{t=1}^{T} \big( y_t - \sum_{k=1}^{p} \beta_k x_{kt} \big)^2 + \lambda \sum_{k=1}^{p} \big| \beta_k \big| \bigg\}.
\label{Eq LASSO Estimation Classic}
\end{align}
Here, two preset tuning parameters $ c $ and $ \lambda $ perform the same role that discourages models from involving too many coefficients in Equation~\eqref{Eq LASSO Estimation Constraint} and Equation~\eqref{Eq LASSO Estimation Classic}, respectively. By shrinking trivial coefficients to zeros, such penalty mechanisms balance residual errors and coefficient numbers and provide an automatic approach for the variable selection.

Instead of directly taking the whole system into account, we refer to the `equation by equation' treatment in \citet{kock2015oracle} to apply LASSO into each equation in Equation~\eqref{Eq Time-Zone VAR Model} separately. Taking the first Asian country as an example, the estimation is presented as follows:
\begin{align}
\widehat{\beta}_1 = \argmin_{\beta_{11},\cdots,\beta_{1N}} \bigg\{ \frac{1}{2} \sum_{t=2}^{T} \big( r_{1t} - \sum_{i \,\in\, As} \beta_{1i} \, r_{1, t-1} - \sum_{j \,\in\, Eu} \beta_{1j} \, r_{j, t-1}  - \sum_{k \,\in\, Am} \beta_{1k} \, r_{k, t-1} \big)^2 + \lambda_1 \sum_{l=1}^N \big| \beta_{1l} \big| \bigg\},
\label{Eq VAR Models with LASSO-First Country}
\end{align}
where $ \lambda_1 $ is a tuning parameter to select those most influential countries for the first one.

%===================================================================================%
%===================================================================================%
%===================================================================================%
%===================================================================================%
%===================================================================================%

\subsection{Model Selections in Time-Zone VAR Models with LASSO}

So far, the tuning parameters $ \bm{\lambda} = \{\lambda_1,\lambda_2, \cdots, \lambda_N \} $ for the equations have been preset before estimation. However, these parameters determine the coefficients in Equation~\eqref{Eq Time-Zone VAR Model}. Further, choosing the optimal $ \bm{\lambda} $ is vital for estimating coefficients in the VAR model and building networks for the global equity market. In this paper, the widely used method, $K$-fold CV, is adopted equation by equation to search the optimal tuning parameters.

Next, take the first country as an example to demonstrate the CV procedure in detail. The full sample is randomly divided into $K$ mutual exclusive folds $ \{ F_1, F_2, \cdots, F_{K} \} $ with roughly equal observations. Then, let $ \Lambda_1 $ be a set of candidate values of $ \lambda_1 $. For the $ k^{th} $ fold ($ k=1,2,\cdots,K $) and $ \lambda \in \Lambda_1 $, let
\begin{align}
\widehat{\beta}_1^{(-k)}(\lambda) = \argmin_{\beta_{11},\cdots,\beta_{1N}} \bigg\{ \frac{1}{2} \sum_{t \notin F_k} \big( r_{1t} - \sum_{l = 1}^N \beta_{1l} \, r_{l, t-1} \big)^2 + \lambda \sum_{l=1}^N \big| \beta_{1l} \big| \bigg\}
\label{Eq VAR Models with LASSO-Cross Validations}
\end{align}
be the LASSO estimator corresponding to  all observations except for those in $ F_k $. Then, the optimal $ \lambda_1 $ is decided by
\begin{align}
\widehat{\lambda}_1 = \argmin_{\lambda \in \Lambda_1} \frac{1}{T} \sum_{k=1}^{K} \sum_{t \in F_k} \big( r_{1t} - \sum_{l = 1}^N \widehat{\beta}_{1l}^{-k} (\lambda) \, r_{l, t-1} \big)^2.
\label{Eq VAR Models with LASSO-Optimal Lambdas}
\end{align}
It is important to emphasize that independent observation is essential for classic CVs to ensure that training pairs are exchangeable, and this assumption may be violated in many settings of financial time series. However, it is analogous to the proof in \citet{bergmeir2018note} to show that the classic CV method still works under the specification Equation~\eqref{Eq Time-Zone VAR Model}. The number of folds $ K $ is selected such that the size of folds is above $ 30 $ (i.e., $ K = \left\lfloor T / 30 \right\rfloor $).\footnote{$ \left\lfloor x \right\rfloor $ is the floor function that returns the greatest integer less than or equal to $ x $.} Besides, since the `equation by equation' treatment is used to estimate VAR coefficients, different equations may have significantly distinct optimal tuning parameters. To minimize the prediction errors in Equation~\eqref{Eq VAR Models with LASSO-Optimal Lambdas}, $ N $ optimal tuning parameters rather than a uniform parameter are used for $ N $ equations.

%===================================================================================%
%===================================================================================%
%===================================================================================%
%===================================================================================%
%===================================================================================%

\subsection{Generating Financial Networks}

For the $K$-fold CV procedure, the random partitions of all observations are of great significance when selecting optimal tuning parameters in Equation~\eqref{Eq VAR Models with LASSO-Optimal Lambdas}, which may eventually cause the huge difference in coefficients in Equation~\eqref{Eq Time-Zone VAR Model}. Thus, a new treatment is implemented to minimize the randomness influence of classic K-fold CV methods. For example, to search the optimal $ \lambda_1 $ in Equation~\eqref{Eq VAR Models with LASSO-First Country}, we regenerate $ R $ times K folds for the full sample and employ Equation~\eqref{Eq VAR Models with LASSO-Optimal Lambdas} to choose the optimal $ \widehat{\lambda}_{1,r} $ in the $ r^{th} $ time.

To generate the adjacency matrix, we consider the $ M $ most frequent tuning parameters $ (\widehat{\lambda}_1^{(1)}, \widehat{\lambda}_1^{(2)}, $  $ \cdots, \widehat{\lambda}_1^{(M)}) $ among $ R $ replications with frequencies $ (Q_1(1), Q_1(2), \cdots, Q_1(M)) $, ranking from largest to smallest, and select the optimal value as
\begin{align}
\widehat{\lambda}_1^{*} = \max \big( \widehat{\lambda}_1^{(1)}, \widehat{\lambda}_1^{(2)}, \cdots, \widehat{\lambda}_1^{(M)} \big).
\label{Eq Optimal Lambdas for Adjacency Matrices}
\end{align}
Then, the adjacency matrix $ \bm{A} = \{ a_{ij} \} $ is defined as
\begin{align}
a_{1j} = \left\lbrace \begin{array}{cc}
	\widehat{\beta}_{j1} (\widehat{\lambda}_1^{*}) / |\widehat{\beta}_{j1} (\widehat{\lambda}_1^{*})|,   &   \widehat{\beta}_{j1} (\widehat{\lambda}_1^{*}) \neq 0,   \\
	0,                                  &   \widehat{\beta}_{j1} (\widehat{\lambda}_1^{*})   =  0.
\end{array} \right.
\label{Eq Adjacency Matrices}
\end{align}
The LASSO regularity path \citep{efron2004least} ensures that the link set decided by $ \widehat{\lambda}_1^* $ is the maximum subset of link sets determined by $ \widehat{\lambda}_1^{(1)}, \widehat{\lambda}_1^{(2)}, \cdots, \widehat{\lambda}_1^{(M)} $.

To construct the weight matrix $ \bm{W} = \{ w_{ij} \} $, we consider coefficients estimated by $ M $ tuning parameters and weight them as
\begin{align}
w_{1j} = |a_{1j}| \times \sum_{m=1}^M \frac{Q_1(m)}{\sum_{n=1}^M Q_1(n)} \widehat{\beta}_{j1} (\widehat{\lambda}_1^{(m)}).
\label{Eq Weight Matrices 1}
\end{align}
Since $ \widehat{\lambda}_1^* $ is not always the most frequent one in $ M $ tuning parameters, we adopt the weighted coefficients instead of $ \widehat{\lambda}_1^* $ to take into account information in those most frequent tuning parameters. The first term $ |a_{1j}| $ in Equation~\eqref{Eq Weight Matrices 1} guarantees that created weighted matrixes coincide with adjacency matrixes. The empirical data is used in Section \ref{Section Comparisons on Network Structures} to demonstrate the advantage of the improved CV method in generating stable network structures.
% \textcolor{red}{Networks generated by Equation~\eqref{Eq Adjacency Matrices} can be found in the Appendix. Figure ? compares our CV procedure with the traditional one when creating financial networks.}

%===================================================================================%
%===================================================================================%
%===================================================================================%
%===================================================================================%
%===================================================================================%

\subsection{Sample Divisions}
\label{Appendix A sample divisions}

The subprime crisis is one of the largest global crises since the Great Depression. While the crisis initially originated in United States in a relatively small segment of the lending market, namely the subprime mortgage market, it rapidly spread across all advanced and emerging economies and economic sectors. It also affected equity markets worldwide, with many countries experiencing even sharper equity market crashes than United States. The European debt crisis was when several European countries experienced the collapse of financial institutions, high government debt, and rapidly rising bond yield spreads in government securities. We divide the sample period into five-time intervals: (1) before the subprime crisis (1 August 2006 to 31 July 2007), (2) during the subprime crisis (1 August 2007 to 31 March 2009), and (3) after the subprime crisis (1 April 2009 to 30 November 2009), (4) during the European debt crisis (1 December 2009 to 16 December 2013), and (5) after the European debt crisis (17 December 2013 to 31 December 2015).

The intervals were determined using landmark events, equity markets' volatilities, and economic growth. August 2007 is usually regarded as the beginning of the subprime mortgage crisis because this month witnessed a series of profound events that eventually contributed to the global stock market plunges. Those landmark events include the German bank, IKB Deutsche Industriebank, announcing a profit warning on 2 August due to its Rhineland Funding; the tenth-largest mortgage lender in United States filing for bankruptcy protection in the court on 6 August; Bear Stearns, the fifth-largest investment bank in United States, announcing the closure of its two funds on 8 August; and France's largest bank, BNP Paribas, announcing the freezing of its three funds on 9 August. The United States GDP rose in the second quarter of 2009 and maintained this quarterly growth for the following ten years. Also, the non-agricultural employment rate continuously increased from the second quarter of 2009. These growths indicate that the negative impact of the subprime crisis was diminishing, and the United States economy was recovering from this crisis. In the meantime, the European debt crisis erupted in the wake of the subprime crisis around late 2009. It was characterized by an environment of overly high government structural deficits and accelerating debt levels.

The European debt crisis, spanning from late 2009 to 2015, was a considerable financial upheaval that primarily affected the Eurozone countries. This crisis was instigated by a confluence of factors, including elevated government debt levels, structural economic frailties, and the subprime crisis in 2008. Greece, Portugal, Ireland, Spain, and Italy were among the countries most heavily impacted, as they grappled with soaring budget deficits and unsustainable debt levels.

The European debt crisis outbreak phase commenced in December 2009 with the emergence of the Greek debt crisis, which subsequently spread to Portugal and Ireland. In November 2010, Ireland succumbed to pressure and accepted assistance, preventing the crisis from intensifying further. Subsequently, the European Central Bank raised interest rates in April and July 2011, leading to the crisis spreading to Italy and Spain. This development also impacted core economies such as France and Germany.

Ireland became the first country to exit the European debt relief mechanism on 16 December 2013, signifying the beginning of the post-European debt crisis period. In May 2014, Portugal exited the bailout mechanism. In 13 July 2015, Greece reached an agreement with the European Union to secure a third round of aid in exchange for implementing reforms. Prime Minister Alexis Tsipras promoted a series of reforms aimed at alleviating the burden on businesses and the public, ultimately putting the Greek economy back on track. Since 2016, the Greek economy has experienced growth.

\subsection{Difficulties of Applying Static Estimation to Study COVID-19 Crisis}
\label{Appendix A Not consider static estimation for COVID-19 crisis}

We do not apply the static estimation approach to study the global equity market before, during, and after the COVID-19 crisis for the following three reasons:

\paragraph{Different COVID-19 severity.} We apply the static estimation method to study the global crisis with bear and bull market transformation characteristics. Therefore, the subprime crisis and the European debt crisis caused by the distrust of the market and quickly spread in the global financial market have a unified series of landmark events that can help researchers establish breaking points (see Appendix \ref{Appendix A sample divisions} for details on those landmark events). The COVID-19 crisis differs from the subprime and European debt crises. First, the start time is difficult to determine. The impact of the COVID-19 crisis on the stock market is an obvious external shock related to the severity of the local pandemic. Since the outbreaks did not occur in all countries at the same time, the start time of the COVID-19 crisis on the financial market in each country is different. For example, when the pandemic broke out in Asia in January 2020, stocks fell sharply; after the outbreak of the pandemic in Europe and Americas at the end of February 2020, it led to multiple circuit breakers in United States. Second, the end time is difficult to determine. Upon the outbreak of the COVID-19 crisis, major central banks worldwide adopted monetary policies that caused the European and American stock markets to fall sharply in a short period and generally rebounded by a large margin after April 2020. More importantly, the COVID-19 crisis poses a heterogeneous impact on economies and corporates \citep{augustin2022sickness, ding2021corporate, hasan2023covid, duchin2021covid}, leading to the very different end time of the COVID-19 crisis in each country and making it unreasonable to identify a unified time division. Besides, a rapidly growing body of research treats COVID-19 as a temporal and external shock rather than a long-term and endogenous crisis, making the applicant of static analysis to the COVID-19 period less plausible. Notably, the COVID-19 pandemic gives rise to a distributional shock to asset prices and aggregate demand \citep{caballero2021model}, sovereign credit risk \citep{augustin2022sickness}, and capital allocation \citep{duchin2021covid}. To match the empirical evidence, an influential body of work highlights the importance of modeling COVID-19 as a productivity shock on firms \citep{barry2022corporate, guerrieri2022macroeconomic, eichenbaum2021macroeconomics, acemoglu2021optimal, baqaee2022supply}.

\paragraph{Short duration of COVID-19.} The subprime crisis lasted over a year, and the European debt crisis lasted about two years. It is hard to say whether the impact of the COVID-19 crisis is a long-term bear market, a bull market, or a transitional reaction to unknown events and subsequent stock market corrections caused by monetary policies. For instance, the COVID-19 crisis is the shortest United States recession on record \citep{berger2021banking}. The United States stock market was in a bear market from February to early March. After that, the Federal Reserve issued the quantitative easing policy in the middle of March, and the United States stock market began to rebound to the highest point in history by the end of 2021. Limited by the short duration of the COVID-19 crisis, using the static method may suffer from a severe data shortage problem and result in significant under-fitting biases. Therefore, the static estimation approach requires a larger sample size to produce valid and robust results.

\paragraph{Non-stationary exogenous variables to describe COVID-19 infections.} In static analysis, considering exogenous variables on COVID-19 infections (e.g., daily new cases or daily deaths) is vital to characterizing the external shock from the COVID-19 pandemic on the financial system and its heterogeneity effect on different national equity markets. This is because the COVID-19 crisis derives from an unprecedented global pandemic, fundamentally different from the subprime and European debt crises. Fears about the deadly virus create tremendous stress on the financial markets and cause massive panic in stock markets, forcing governments to announce a series of economic policies to stabilize markets. However, these COVID-19 exogenous variables are typically non-stationary time series and cannot be directly incorporated into equations under the VAR specification, making statistical inference more difficult in static estimation.

\smallskip
Given the above reasons, we adopt the dynamic approach using the rolling estimation method to study the contagion mechanism in the global equity market over a long period and examine the continuous real-time evolution of systemwide connectedness. Although the dynamic method still suffers from the lack of COVID-19 exogenous variables, it can overcome the difficulty of dividing sample periods and the problem of insufficient sample size in static analysis.

%===================================================================================%
%===================================================================================%
%===================================================================================%
%===================================================================================%
%===================================================================================%

\subsection{Reasons of Excluding Explanatory Variables in Equation \eqref{Eq Time-Zone VAR Model}}
\label{Appendix A Reasons of Excluding Explanatory Variables in Equation 1}

We argue that our econometric framework defined in Equation \eqref{Eq Time-Zone VAR Model} can identify the mechanism of financial contagion without involving explanatory or observable variables because of theoretical and empirical reasons.

\paragraph{Theoretical view.} VAR models are widely used to investigate the contagion mechanisms of financial markets \citep[e.g.,][]{billio2012econometric, diebold2009measuring, diebold2011on}. Specifically, VAR coefficients can reveal the modifications of the transmission mechanism \citep{claeys2014measuring}.

To the best of our knowledge, nearly all literature on VAR networks does not consider explanatory variables and directly study the contagion mechanism \citep[e.g.,][]{billio2012econometric, diebold2009measuring, diebold2011on, demirer2018estimating}. As the most related work, \citet{billio2012econometric} use the Granger-causality test and monthly return data to study the network of financial institutions from four sectors without considering any explanatory variable.

One of two exception papers from \citet{alter2014dynamics} introduces daily explanatory variables to capture common global and regional factors during the European debt crisis. However, our framework is entirely different because selected national equity markets are either irreplaceable in the global system (global factors) or representative enough to reflect regional systems (regional factors). In the other work, \citet{claeys2014measuring} use the factor analysis technique to extract common factors and then use the VAR decomposition to study EU sovereign bond markets. Given the same reason, our framework involves enough distributed and significant national equity markets to reflect the common factors. Moreover, we follow \citet{claeys2014measuring} to extract common factors before estimating VAR coefficients, and the relevant results are consistent with our previous findings.

\paragraph{Empirical view.} The lack of daily frequency data is the main failure reason for incorporating explanatory variables into our framework in practice. Classic explanatory variables like industry or macroeconomic factors are reported quarterly or annually, and hence, these variables are not available at a daily frequency. More importantly, explanatory variables like industry or macroeconomic factors may not be readily applied in our model based on intraday data \citep{demirer2018estimating} and cause the severe mixed frequency problem \citep{GHYSELS200659}. Besides, countries may also differ in the definition of industry and macroeconomic factors, and including such factors may complicate the economic interpretation of our model \citep{rapach2013international}.

% We also admit more suitable explanatory variables or sophisticated econometric models may improve the validation of our framework and leave it for future work.

%===================================================================================%
%===================================================================================%
%===================================================================================%
%===================================================================================%
%===================================================================================%

%===================================================================================%
%===================================================================================%
%===================================================================================%
%===================================================================================%
%===================================================================================%

%===================================================================================%
%===================================================================================%
%===================================================================================%
%===================================================================================%
%===================================================================================%

\clearpage

%		Numeric results on in-sample ratios are presented in Table \ref{Tab In-sample ratios R2} in Appendix \ref{Appendix B}.

\clearpage
\section{Comparison between Classic and Improved CV Methods}\label{Section Comparisons on Network Structures}

We propose an improved CV procedure to generate networks with stable structures. The classic VAR model may suffer from the high-dimensional problem and produce bad fitting results because the agent number ($N$) is large but the observation number ($T$) is limited.\footnote{The high-dimensional problem refers to a situation where unknown parameters have a comparable size to observations. When building networks by the VAR specification, we need to use $N \times T$ observations to estimate $N \times N$ unknown parameters.} The LASSO-type estimate provides an appealing solution to the high-dimensional problem \citep{demirer2018estimating}. However, the LASSO-type estimate relies on the classic CV procedure to select the optimal tuning parameter, during which random partitions may create networks with unignorable structure differences in replications even using the same data. Consequently, an improved CV method is developed to moderate the problem of instability caused by random partitions to produce stable networks.

We use the empirical data to demonstrate the differences in generating LASSO networks via the classic $K$-fold CV and the improved CV in Equation~\eqref{Eq Optimal Lambdas for Adjacency Matrices}.\footnote{We consider the top $ 5 $ most frequent tuning parameters in each replication ($ M=5 $).} Independently producing networks $ 300 $ times, we consider the network density and the mutual proportion to mirror the stability of networks. The network density is a ratio of the existing link number to the maximum link number in theories\footnote{The maximum link number is equal to $ N(N-1)/2 $ and $ N $ is the number of national stock markets.}, and the mutual proportion is a ratio of the number of mutual links shared by the first $ r $ networks ($ r $ is the replicate time) to the maximum link number.

\begin{landscape}
\begin{figure}
\begin{center}
	%	\makebox[\textwidth]{\includegraphics[width=\textwidth]{Figures/CampareImprovedClassicCV.eps}}
	\includegraphics[width=\linewidth]{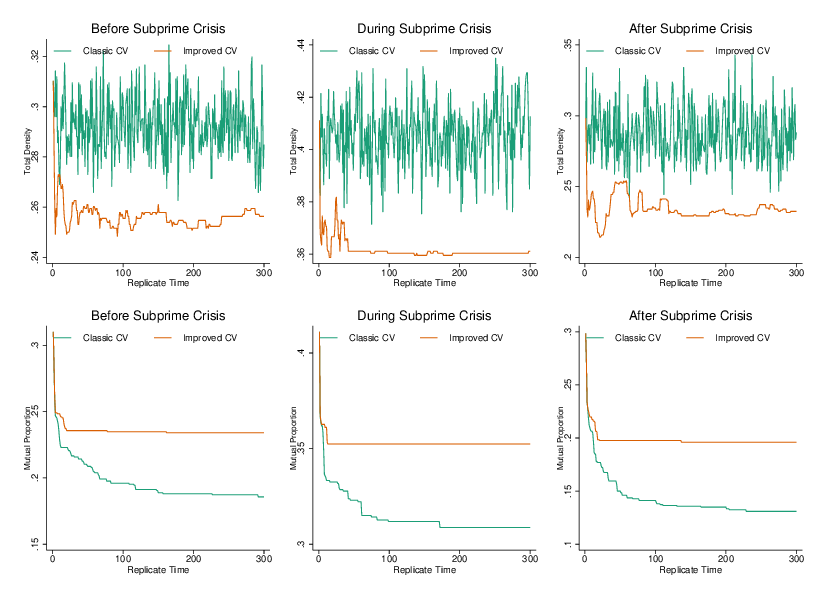}
	\caption{\footnotesize \textbf{Densities and mutual proportions of networks generated by the classic CV and the improved CV over the different periods.} Orange and green solid lines represent classic CV networks and improved CV networks, respectively. The network density is a ratio of the existing link number to the maximum link number in theories, and the mutual proportion is a ratio of the number of mutual links shared by the first $ r $ networks ($ r $ is the replicate time) to the maximum link number.}
	\label{Fig Stable Network Stuctures 1}
\end{center}	
\end{figure}
\end{landscape}

\begin{figure}
\begin{center}
	%	\makebox[\textwidth]{\includegraphics[width=\textwidth]{Figures/CampareImprovedClassicCV.eps}}
	\includegraphics[width=\textwidth]{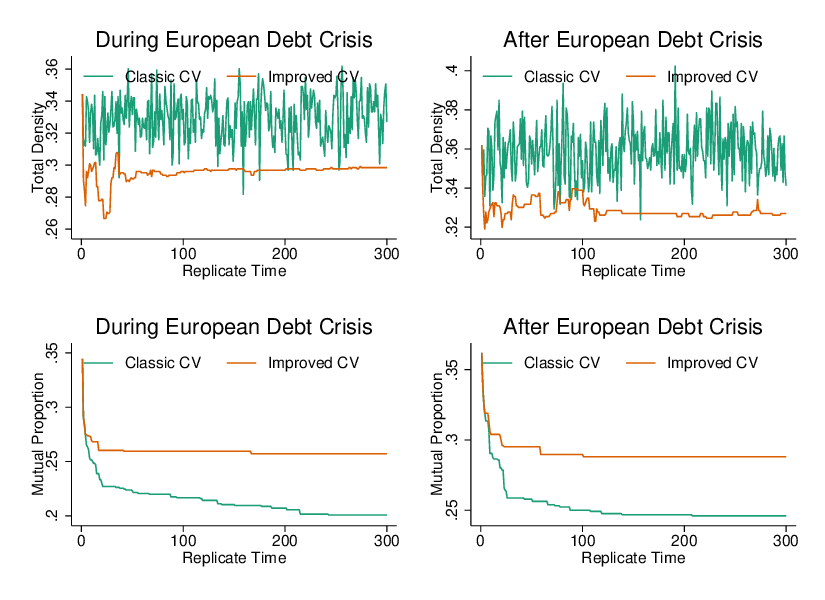}
	\caption{\textbf{Densities and mutual proportions of networks generated by the classic CV and the improved CV over the different periods.} Orange and green solid lines represent classic CV networks and improved CV networks, respectively. The network density is a ratio of the existing link number to the maximum link number in theories, and the mutual proportion is a ratio of the number of mutual links shared by the first $ r $ networks ($ r $ is the replicate time) to the maximum link number.}
	\label{Fig Stable Network Stuctures 2}
\end{center}	
\end{figure}

Figures \ref{Fig Stable Network Stuctures 1} and \ref{Fig Stable Network Stuctures 2} present densities and mutual proportions of LASSO-type networks generated by the classic CV (solid orange lines) and by the improved CV (solid green lines) over three periods. As the upper panel shows, densities in classic CV networks display dramatic fluctuations among 300 replications, indicating that the classic CV fails to generate a stable network structure. By contrast, the improved CV method alleviates this instability problem and keeps the density varying within a small range. The bottom panel shows the mutual proportion in the classic CV network falls considerably as the replicate time rises, demonstrating the evident differences of the network structure created by the classic CV method in each replication. As a comparison, the improved CV method notably increases the mutual proportion and fixes this ratio when the replication time is larger than 50.

\clearpage
\newgeometry{margin=1cm,bmargin=1in,tmargin=1in}
\restoregeometry

\clearpage

%===================================================================================%
%===================================================================================%
%===================================================================================%
%===================================================================================%
%===================================================================================%

\clearpage
\section{VAR Coefficients and Link Signs}
\label{Appendix VAR Coefficients and Link Signs}

\begin{table}[!htbp]
\begin{center}
\begin{spacing}{1}

\caption{\textbf{Autoregressive coefficients in the time-zone VAR model.}}
\label{Tab AR Coefficients}
\begin{tabular}{lccccc}
	\toprule
	& \multicolumn{3}{c}{Subprime Crisis} & \multicolumn{2}{c}{Europe Debt Crisis} \\ \cmidrule(lr){2-4} \cmidrule(lr){5-6}
	National Market   & Before      & During     & After      & During         & After          \\ \midrule
	Australia   & -0.146      & -0.141     & 0      & -0.080         & 0          \\
	Malaysia    & 0       & -0.038     & 0      & 0.088          & 0.055          \\
	Indonesia   & 0       & 0.023      & 0      & 0          & 0          \\
	Korea       & 0       & 0      & -0.129     & -0.149         & 0          \\
	Japan       & -0.023      & -0.149     & 0      & 0          & -0.151         \\
	Singapore   & 0       & 0      & 0      & -0.051         & 0          \\
	NewZealand  & 0       & 0      & 0      & 0          & 0.016          \\
	Philippines & 0       & 0.052      & 0      & 0.073          & 0          \\
	Thailand    & -0.140      & -0.048     & 0      & 0          & 0          \\
	China       & 0       & 0      & 0      & 0          & 0          \\
	Hong Kong   & -0.096      & -0.215     & 0      & 0          & 0          \\ \midrule
	Netherlands & 0       & 0      & 0      & 0          & 0          \\
	Greece      & 0       & 0      & 0      & 0.017          & 0.040          \\
	Belgium     & 0       & 0      & 0      & -0.027         & 0          \\
	France      & 0       & -0.141     & 0      & 0          & 0          \\
	Germany     & 0       & 0      & 0      & 0          & 0          \\
	Finland     & 0       & 0      & 0      & 0          & 0          \\
	Spain       & 0       & 0      & 0      & 0          & -0.019         \\
	Ireland     & 0       & 0      & 0      & 0          & 0          \\
	Italy       & 0       & 0      & 0      & 0          & -0.051         \\
	Denmark     & 0       & 0      & 0      & 0          & 0          \\
	Norway      & -0.025      & 0      & 0      & 0          & -0.097         \\
	Sweden      & 0       & 0      & 0      & 0          & -0.113         \\
	Portugal    & 0       & 0      & 0      & 0.097          & 0.044          \\
	Russia      & 0       & 0.039      & 0      & 0          & 0          \\
	Switzerland & 0       & 0      & -0.006     & 0          & 0          \\
	United Kingdom          & 0       & 0      & 0      & 0          & 0          \\
	Poland      & 0       & 0      & 0.051      & 0          & 0          \\
	Turkey      & -0.037      & 0      & 0      & 0          & -0.034         \\
	Austria     & 0       & -0.034     & -0.192     & 0          & 0          \\ \midrule
	Brazil      & -0.049      & -0.074     & -0.028     & 0          & -0.049         \\
	Chile       & 0.065       & 0.060      & 0      & 0.193          & 0.032          \\
	Argentina   & -0.101      & 0      & 0      & 0.024          & 0          \\
	Mexico      & 0       & 0      & 0      & -0.011         & 0          \\
	Canada      & -0.047      & -0.163     & 0      & 0          & 0          \\
	United States   & -0.123      & -0.286     & -0.067     & -0.097         & -0.060        \\ \bottomrule
\end{tabular}

\end{spacing}
\end{center}
\end{table}

\clearpage
\begin{landscape}

\begin{figure}
\begin{center}
	\includegraphics[scale=0.9]{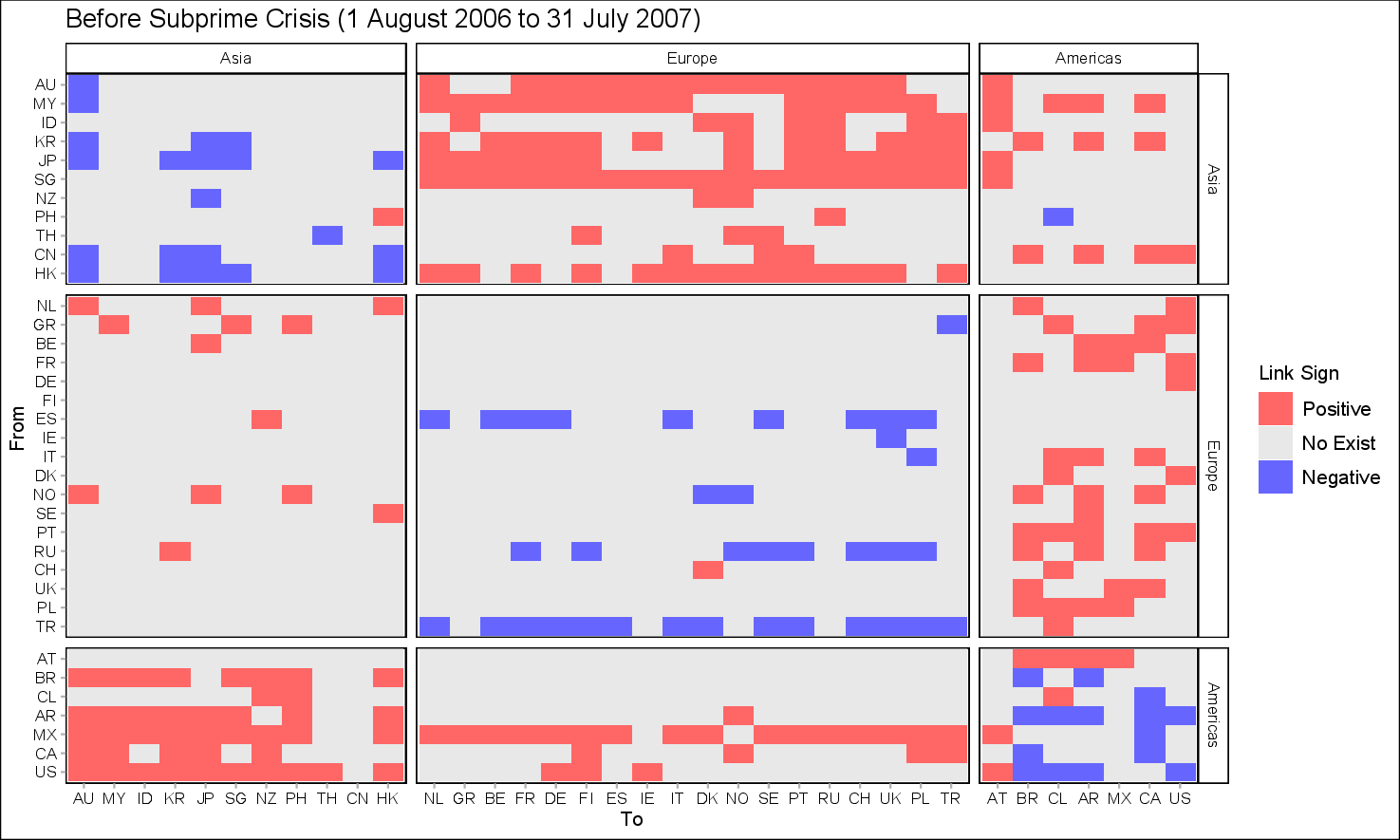}
		
	\caption{\textbf{The heatmap of link signs in the generated global network before the subprime crisis.} Light grey blocks represent the corresponding links do not exist, and red and blue blocks represent positive and negative link signs, respective, decided by Equation \eqref{Eq Adjacency Matrices}.}
	
	\label{Fig HeatMapVARCoefs_SubprimeBefore}
\end{center}
\end{figure}

\begin{figure} 
\begin{center}
	\includegraphics[scale=0.9]{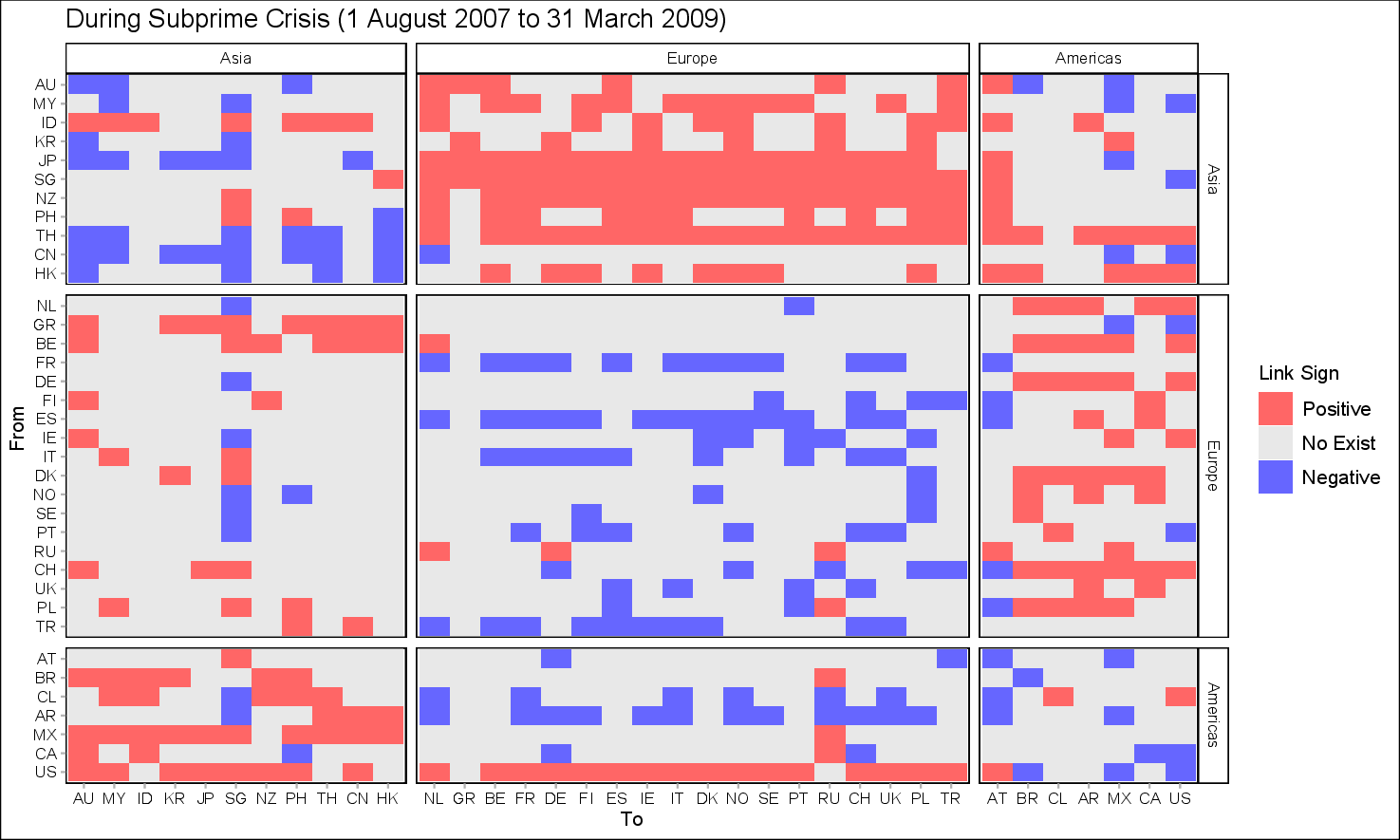}
	
	\caption{\textbf{The heatmap of link signs in the generated global network during the subprime crisis.} Light grey blocks represent the corresponding links do not exist, and red and blue blocks represent positive and negative link signs, respective, decided by Equation \eqref{Eq Adjacency Matrices}.}
	
	\label{Fig HeatMapVARCoefs_SubprimeDuring}
\end{center}
\end{figure}

\begin{figure} 
\begin{center}
	\includegraphics[scale=0.9]{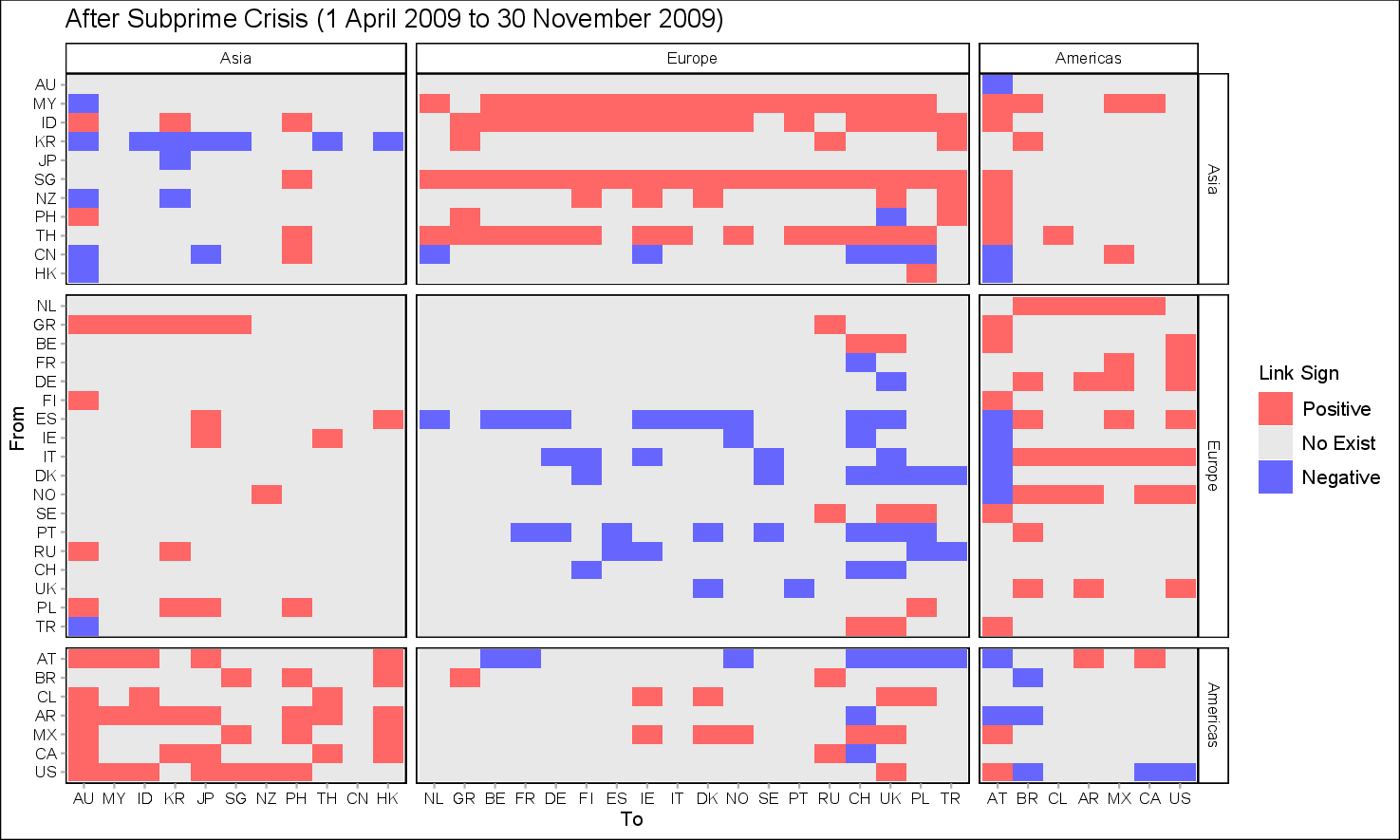}
	
	\caption{\textbf{The heatmap of link signs in the generated global network after the subprime crisis.} Light grey blocks represent the corresponding links do not exist, and red and blue blocks represent positive and negative link signs, respective, decided by Equation \eqref{Eq Adjacency Matrices}.}
	
	\label{Fig HeatMapVARCoefs_SubprimeAfter}
\end{center}
\end{figure}

\begin{figure} 
\begin{center}
	\includegraphics[scale=0.9]{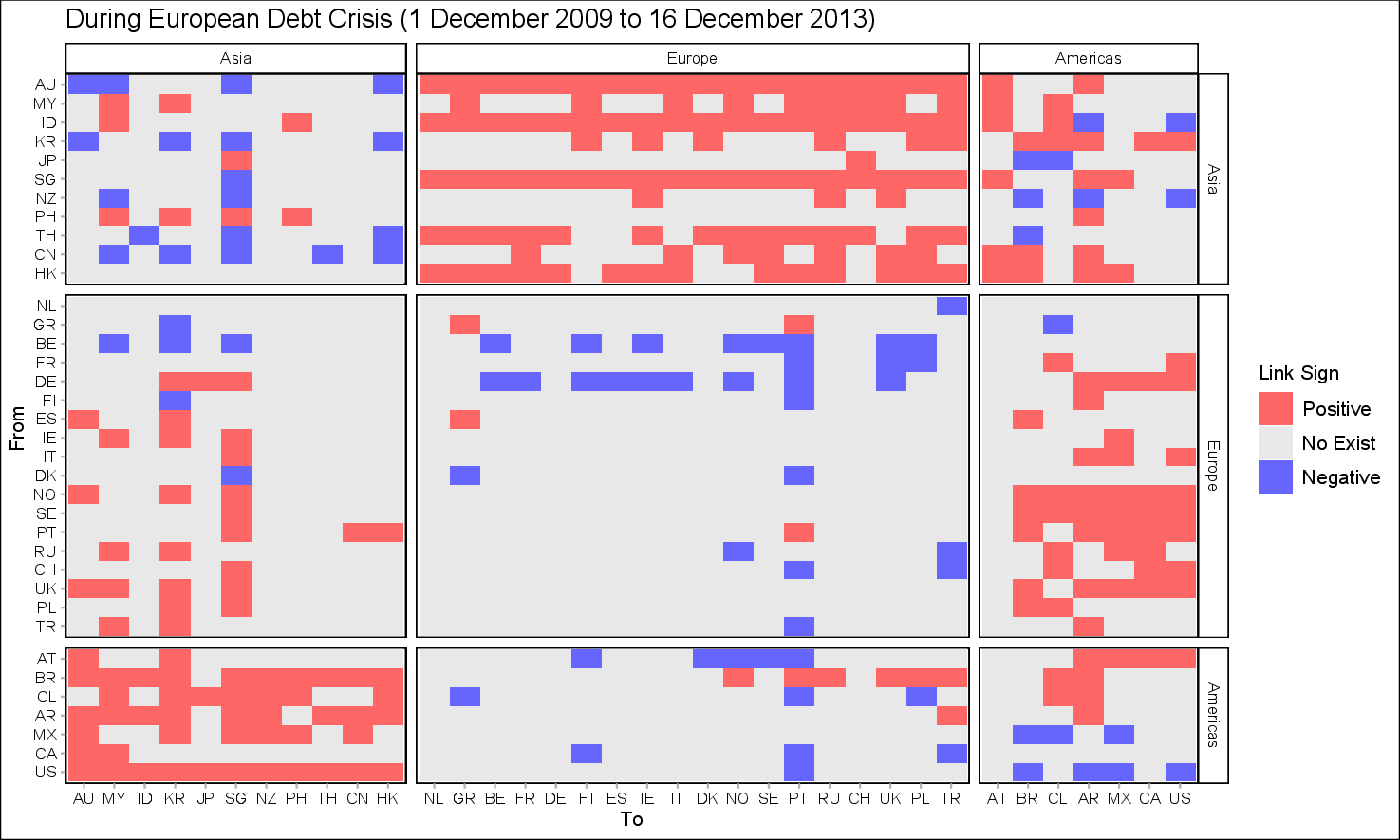}
	
	\caption{\textbf{The heatmap of link signs in the generated global network during the European debt crisis.} Light grey blocks represent the corresponding links do not exist, and red and blue blocks represent positive and negative link signs, respective, decided by Equation \eqref{Eq Adjacency Matrices}.}
	
	\label{Fig HeatMapVARCoefs_EuDebtDuring}
\end{center}
\end{figure}

\begin{figure} 
\begin{center}
	\includegraphics[scale=0.9]{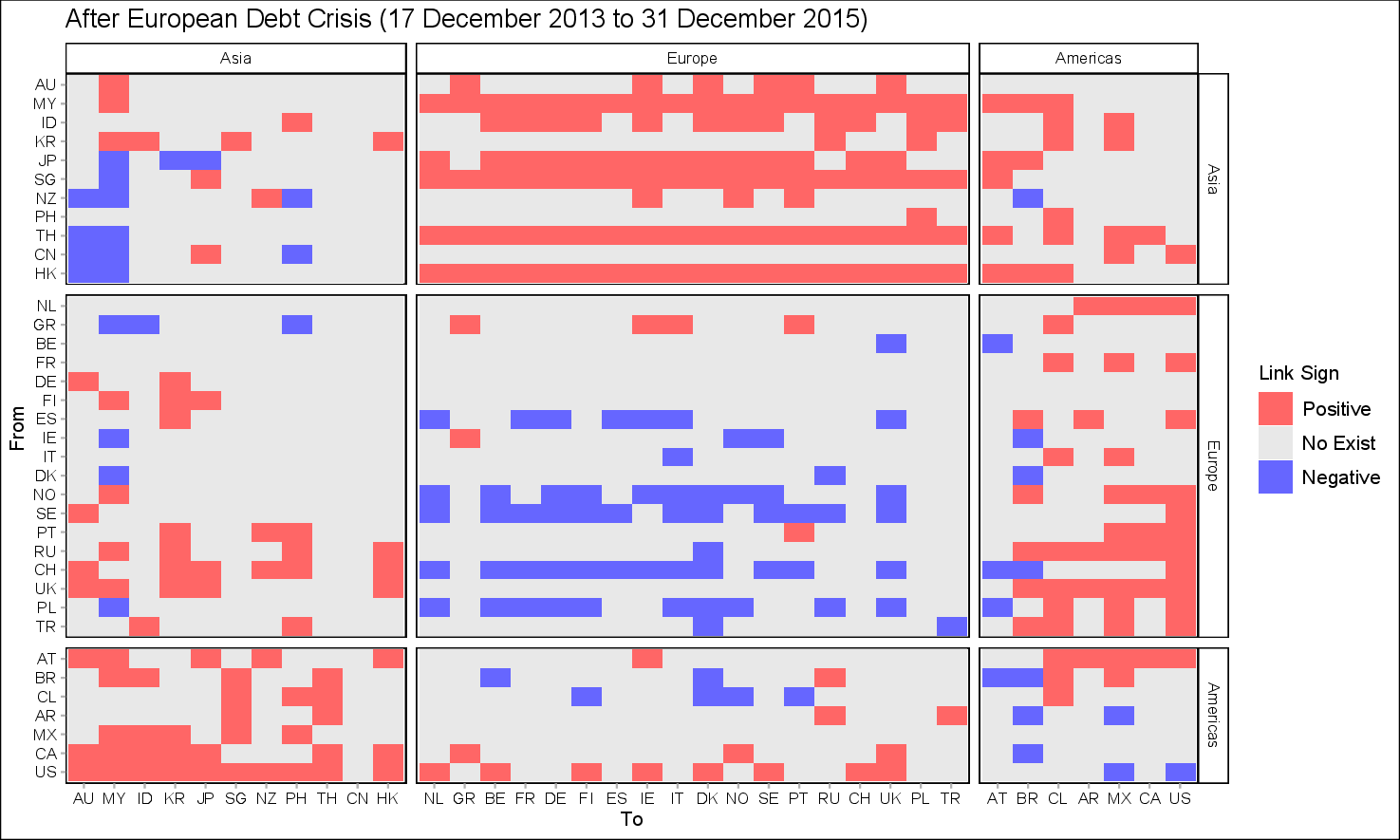}
	
	\caption{\textbf{The heatmap of link signs in the generated global network after the European debt crisis.} Light grey blocks represent the corresponding links do not exist, and red and blue blocks represent positive and negative link signs, respective, decided by Equation \eqref{Eq Adjacency Matrices}.}
	
	\label{Fig HeatMapVARCoefs_EuDebtAfter}
\end{center}
\end{figure}

\end{landscape}

%===================================================================================%
%===================================================================================%
%===================================================================================%
%===================================================================================%
%===================================================================================%

%
%\newpage
%\section{Dynamic Results}
%\label{AppendixDynamic Results}

%\end{CJK}
\end{document}